\pdfoutput=1
\documentclass[aps,prd,amsmath,floats,floatfix,twocolumn,superscriptaddress,nofootinbib,showpacs]{revtex4-2}

\usepackage[T1]{fontenc}
\usepackage[utf8]{inputenc}
\usepackage{lmodern}
\usepackage{lipsum}

\usepackage[dvipsnames, usenames]{xcolor}
\definecolor{linkcolor}{rgb}{0.0,0.3,0.5}
\usepackage[hypertexnames=false, unicode, colorlinks=true, linkcolor=linkcolor,
citecolor=linkcolor, filecolor=linkcolor,urlcolor=linkcolor,
pdfusetitle]{hyperref}

\usepackage[all]{hypcap}
\usepackage{graphicx}
\usepackage{xspace}
\usepackage{amssymb}
\usepackage[normalem]{ulem} 
\usepackage{bm} 
\usepackage{appendix}

\usepackage{microtype}

\usepackage[english]{babel}
\usepackage{blindtext}

\graphicspath{
  {figs/}
}

\usepackage{tikz}
\usetikzlibrary{shapes,snakes}
\usetikzlibrary{arrows,intersections,patterns,decorations.markings,shapes.geometric}
\usetikzlibrary{calc,fadings,decorations.pathreplacing,positioning}
\usetikzlibrary{decorations.pathmorphing}
\tikzset{snake it/.style={decorate, decoration=snake}}

\usepackage{pgfplots}
\usepgfplotslibrary{colormaps}
\usetikzlibrary{intersections,patterns,pgfplots.fillbetween}

\tikzset{->-/.style={decoration={
  markings,
  mark=at position .5 with {\arrow{>}}},postaction={decorate}}
}
\tikzset{-<-/.style={decoration={
  markings,
  mark=at position .5 with {\arrow{<}}},postaction={decorate}}
}

\tikzset{
  >=latex, 
  inner sep=0pt,
  outer sep=2pt,
  mark coordinate/.style={inner sep=0pt,outer sep=0pt,minimum size=3pt,
    fill=black,circle}
}

\DeclareMathAlphabet{\mathpzc}{OT1}{pzc}{m}{it}

\renewcommand{\vec}[1] {\bm{#1}}
\newcommand{\vhat}[1]{\vec{\hat{#1}}}

\newcommand*{\df}  {\delta}

\newcommand*{\non} {\nonumber}
\newcommand*{\lb} {\left(}
\newcommand*{\rb} {\right)}

\newcommand*{\la} {\left\langle}
\newcommand*{\ra} {\right\rangle}

\newcommand{\eq}[1]{\begin{align}#1\end{align}}
\newcommand{\eeq}[1]{\begin{equation}#1\end{equation}}

\definecolor{darkred}{RGB}{175,0,0}
\definecolor{darkblue}{RGB}{14,0,185}

\begin{document}

\rightline{\scriptsize RBI-ThPhys-2023-14}

\title{The observed power spectrum \& frequency-angular power spectrum}

\newcommand\alvisehome{
\affiliation{Dipartimento di Fisica Galileo Galilei, Universit\` a di Padova, I-35131 Padova, Italy}
\affiliation{INFN Sezione di Padova, I-35131 Padova, Italy}
\affiliation{INAF-Osservatorio Astronomico di Padova, Italy}
\affiliation{Theoretical Physics Department, CERN, 1 Esplanade des Particules, 1211 Geneva 23,
Switzerland}
}
\newcommand\zvonehome{
\affiliation{Ru\dj er Bo\v{s}kovi\'c Institute, Bijeni\v{c}ka cesta 54, 10000 Zagreb, Croatia}
\affiliation{Kavli Institute for Cosmology, University of Cambridge, Cambridge CB3 0HA, UK}
\affiliation{Department of Applied Mathematics and Theoretical Physics, University of Cambridge, Cambridge CB3 0WA, UK }
}

\author{Alvise Raccanelli}
\email{alvise.raccanelli.1@unipd.it}
\alvisehome

\author{Zvonimir Vlah}
\email{zvlah@irb.hr}
\zvonehome

\begin{abstract}
The two-point summary statistics is one of the most commonly used tools in the study of cosmological structure. Starting from the theoretical power spectrum defined in the 3D volume and obtained via the process of ensemble averaging, we establish the construction of the \textit{observed 3D power spectrum}, folding the unequal-time information around the average position into the wave modes along the line of sight. We show how these unequal-time cross-correlation effects give rise to scale-dependent corrections in the observable 3D power spectrum. We also introduce a new dimensionless observable, the \textit{frequency-angular power spectrum}, which is a function of dimensionless and directly observable quantities corresponding to Fourier counterparts of angles and redshifts. While inheriting many useful characteristics of the canonical observed power spectrum, this newly introduced statistic does not depend on physical distances and is hence free of so-called Alcock-Paczy\'nski effects. Such observable thus presents a clear advantage and simplification over the traditional power spectrum.

Moreover, relying on linear theory calculations, we estimate that unequal-time corrections, while generally small, can amount to a few percent on large scales and high redshifts. Interestingly, such corrections depend on the bias of the tracers, the growth rate, but also their time derivatives, opening up the possibility of new tests of cosmological models. These radial mode effects also introduce anisotropies in the observed power spectrum, in addition to the ones arising from redshift-space distortions, generating non-vanishing odd multiples and imaginary contributions. Lastly, we investigate the effects of unequal-time corrections in resumming long displacements (IR-resummation) of the observed power spectrum.
\end{abstract}

\maketitle

\section{Introduction}
\label{sec:intro}

The cosmological large-scale structure offers a competitive and promising avenue for extracting physical information on our Universe from the distribution of matter. Next-generation galaxy surveys, like Euclid~\cite{Amendola:2016}, DESI~\cite{Aghamousa:2016}, Rubin~\cite{Abate:2012}, Roman~\cite{Spergel:2015}, SPHEREx~\cite{Dore:2014}, SKAO~\cite{Bacon++:2018}, MegaMapper \cite{Schlegel:2022}, ATLAS~\cite{Spezzati:23ATLAS} and others, aim to address various cosmological questions, ranging from uncovering the nature of dark energy and tests of general relativity on large scales~\cite{Weinberg:2013, Mortonson:2013}, to constraining the properties of the initial conditions of the universe by measuring signals of primordial non-Gaussianity~\cite{Ross++:2012, Alvarez:2014}. In order to successfully extract accurate cosmological information, robust measurements and reliable statistical tools are of paramount importance. For this purpose, the two-point statistics in Fourier space, be it in 3D (the power spectrum) or in 2D (the angular power spectrum), has been the observable of choice for a wide range of these surveys; alternative statistics are e.g.,~the configuration space two-point function and the spherical-Fourier Bessel (for a (incomplete) list of measurements using different procedures see, e.g.,~\cite{Percival++:2004, Eisenstein++:2005, Cole++:2005, Tegmark++:2006, Padmanabhan:2006, Guzzo++:2008, Percival++:2008, Ho:2008, Reid:2009, Kazin++:2009, Ross:2011, Samushia++:2011, Beutler++:2013, Gil-Marin++:2014, Gil-Marin++:2015, Beutler++:2016, Sanchez++:2016, Ivanov:2019, DAmico:2019, Alam++:2020, Zhao:2020, Chen:2021}).
The primary motivation for the choice of Fourier statistics is the linearity of wave modes on the largest cosmological scales, thus ensuring the independence of theoretical errors, subsequently reflected in the diagonal form of the corresponding covariance matrix.

\begin{table*}
\centering
\begin{tabular}{l l}
\hline
\hline
$\df^{\rm K}_{ij}$ &~ Kronecker symbol \\ [1pt]
$\df^{\rm D} (\vec x)$ &~ Dirac delta function \\ [1pt]
$ W(\chi) $ &~ Window function; related to the specific observable and survey\\ [1pt]
\hline
$\df(\vec x)$ &~ 3D density field of matter or biased tracer \\ [1pt]
$\hat \df(\vec \theta)$ &~ 2D projected filed in the real space coordinates on the sky  \\ [1pt]
\hline
$\mathcal P(\vec k;\, z, z')$ &~ Unequal-time theoretical power spectrum of the 3D density field (unobservable) \\ [1pt]
$P(\vec k;\, z)$ &~ Equal-time observed power spectrum (constructed from observable fields) \\ [1pt]
$\mathbb{C}_{\ell}  (z,z')$ &~ Unequal-time angular power spectrum (in the narrow window function limit) \\ [1pt]
$C_{\ell}$ &~ Projected angular power spectrum (with finite size window functions)  \\ [1pt]
$\widetilde {\mathbb C}(\omega, \ell, \bar z)$ &~ Frequency-angular power spectrum (dimensionless, equivalent to the observed power spectrum) \\ [1pt]
\hline
\hline
\end{tabular}
\caption{Notation used for the most important quantities in this paper.}
\label{tab:notation}
\end{table*}

However, due to the complex nature of relating the actual observable to theoretical predictions in most of the analyses (sometimes also called lightcone effects), in practice, one has to resort to a series of simplifications and auxiliary modelling. One such approximation that we focus on in this paper is related to the fact that for a tracer (e.g.,~galaxies) observed in a redshift bin, a 3D power spectrum necessarily needs to incorporate the unequal-time effects related to the redshift difference of the correlated points. Moreover, even in the case when we observe angular positions as well as redshifts of individual tracers, the observables that can be constructed never correspond to the actual 3D unequal-time power spectrum. The reason for this is that the wave modes along the line-of-sight and unequal-time effects are inevitably mixed up and projected on top of each other (see~\cite{Raccanelli+:2023} for a illustration). Such projection effects are ignored in all contemporary practical applications of the 3D power spectrum, relying on existing approximations and beliefs that the correction will be small and negligible. At any rate, this is a notion that ought to be scrutinized, especially in light of the ever-increasing depth and area coverage of forthcoming surveys.

In this paper, we focus on the ab-initio construction of the two-point observables, taking into account these projection effects. On our path to addressing this matter, we find that the issue at hand naturally fragments into the following set of questions:
\begin{itemize}
\item How to construct the flat-sky approximation to the angular power spectrum taking into account unequal time effects?
\item Fixed observer breaks the translation invariance in the plane parallel approximation. How can we quantify these effects? 
\item How to construct the observed 3D power spectrum from the given projected angular correlation statistics?
\item What are the correction to the observed 3D power spectrum induced by these projections and unequal time effects?
\item Is there an alternative statistic to the 3D power spectrum and angular power spectrum capturing the same information content?
\end{itemize}

This paper is organised as follows. In Section~\ref{sec:power_spectrum} we first re-derive the projected angular power spectrum in terms of the unequal-time theoretical power spectrum in the flat-sky approximation. From there, we derive the observed 3D equal-time power spectrum and introduce the new statistics called frequency-angular power spectrum that is free of so-called Alcock-Paczyn\' ski effects. Once these observable statistics are defined, we consider the corrections generated by the unequal-time effects. This is done in Section~\ref{sec:results}. In the same section, we also consider the unequal-time effects arising due to the long displacement field via the IR resummation mechanism. We close the discussion with our concluding remarks in Section~\ref{sec:conclusion}. Details of our analysis are presented in the series of Appendixes~\ref{appA},~\ref{appB} ,~\ref{app:unequal_time} and~\ref{app:UT_LPT}.

We use Planck cosmology~\cite{Aghanim:2018}, where $\Omega_c h^2 = 0.11933$, $\Omega_b h^2 = 0.02242$, $\Omega_K h^2 = 0$, $h=0.6766$, $n_s=0.9665$, and $\sigma_8 = 0.81027941$. In Table~\ref{tab:notation} we provide a short summary of the key physical quantities featured in the paper.  The linear 3D power spectrum can be obtained using the CAMB~\cite{Lewis:1999} or CLASS~\cite{Lesgourgues:2011} codes.

\section{From the theoretical to the observed power spectrum} 
\label{sec:power_spectrum}

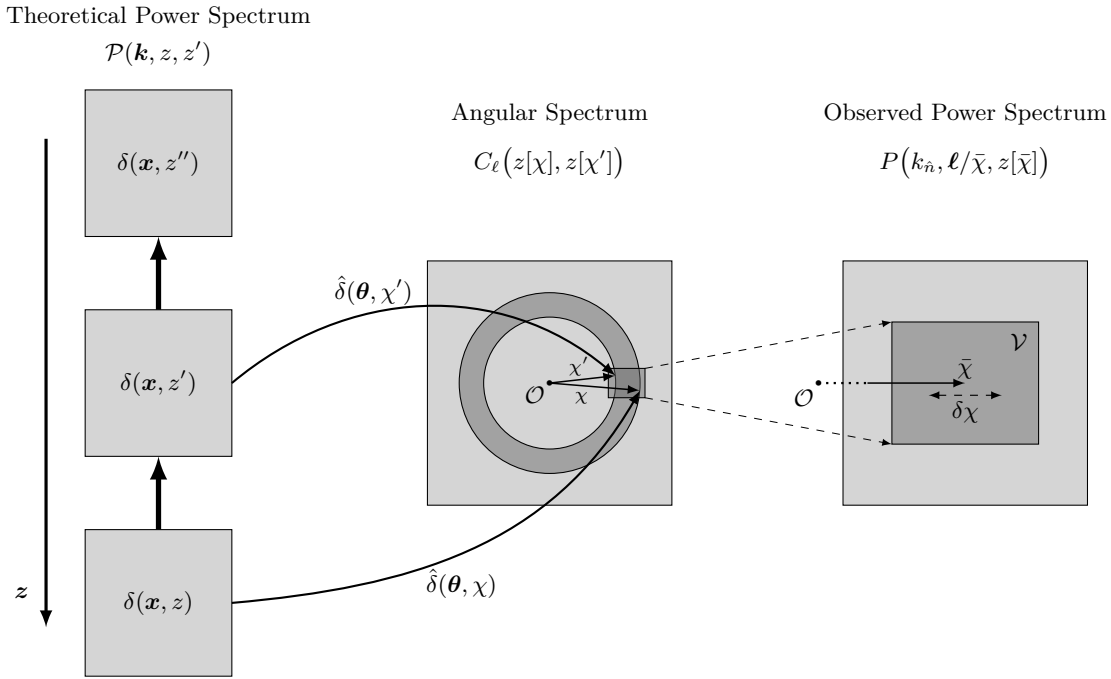
\begin{figure*}[t!]
\centering
\begin{tikzpicture}[scale=.65,every node/.style={minimum size=1cm},on grid]
                  
  \draw[<-, very thick] (-2.3,-5.0) -- (-2.3,5.0);
  \node[black, opacity = 0] at (3,-6) {\textbullet}; 
  
  \draw[->, line width=0.7mm] (0,-3.0) -- (0,-1.5);
  \draw[->, line width=0.7mm] (0,1.5) -- (0,3.0);
  
  \draw[fill=gray!50!white, draw=black, fill opacity = 0.65] (-1.5,3.0) rectangle (1.5, 6.0);
  \draw[fill=gray!50!white, draw=black, fill opacity = 0.65] (-1.5,-1.5) rectangle (1.5,1.5);
  \draw[fill=gray!50!white, draw=black, fill opacity = 0.65] (-1.5,-3.0) rectangle (1.5,-6.0);  

  \draw[fill=gray!50!white, draw=black, fill opacity = 0.65] (5.5,-2.5) rectangle (10.5,2.5);
  \draw[fill=gray!80!white, draw=black, fill opacity = 0.85, even odd rule] (8,0) circle (1.35) (8,0) circle (1.85);
  \node[black] at (8.0,0.0) {\scriptsize \textbullet}; 

  \draw[fill=gray!50!white, draw=black, fill opacity = 0.65] (14.0,-2.5) rectangle (19.0,2.5);

  \draw[fill=gray!80!black, draw=black, fill opacity = 0.45] (9.2,-0.3) rectangle (9.95,0.3);  
  \draw[fill=gray!80!black, draw=black, fill opacity = 0.45] (15.0,-1.25) rectangle (18.0,1.25);

  \draw [->,thick] (1.5,0.0) to [out=40,in=140] (9.35,0.15);
  \draw [->,thick] (1.5,-4.5) to [out=5,in=240] (9.85,-0.15);

  \draw[->, line width=0.2mm] (8.0, 0.0) -- (9.34,0.15);
  \draw[->, line width=0.2mm] (8.0, 0.0) -- (9.84,-0.15);  

  \draw[->, dashed, line width=0.1mm] (9.95,0.3) -- (15.0,1.25);
  \draw[->, dashed, line width=0.1mm] (9.95,-0.3) -- (15.0,-1.25) ;  

  \draw[dotted, line width=0.3mm] (13.5,0) -- (14.5,0);
  \draw[->, line width=0.2mm] (14.5,0) -- (16.5,0);

  \draw[<->, dashed, line width=0.15mm] (15.75,-0.25) -- (17.25,-0.25);

  \node[black] at (13.5,0.0) {\scriptsize \textbullet}; 

  \node[black] at (0,4.5) {\small $\delta(\vec x, z'')$};    
  \node[black] at (0,0) {\small $\delta(\vec x, z')$};
  \node[black] at (0,-4.5) {\small $\delta(\vec x, z)$};
  \node[black] at (6.2,-4.1) {\small $\hat \delta(\bm \theta, \chi)$};  
  \node[black] at (4.4,1.9) {\small $\hat \delta(\bm \theta, \chi')$};  

  \node[black] at (7.7,-0.25) {\small $\mathcal O$};
  \node[black] at (13.2,-0.3) {\small $\mathcal O$};  
  \node[black] at (8.65,-0.3) {\scriptsize $\chi$};
  \node[black] at (8.6,0.35) {\scriptsize $\chi'$};  

  \node[black] at (16.5,0.3) {\small $\bar \chi$};
  \node[black] at (16.5,-0.6) {\small $\delta \chi$};
  \node[black] at (17.6,0.85) {\small $\mathcal V$};

  \node[black] at (-2.8,-4.3) {${\bm z}$};

  \node[black] at (0.0,7.5) {Theoretical Power Spectrum};  
  \node[black] at (0.0,6.75) {\small $\mathcal P (\vec k, z, z')$};
  
  \node[black] at (8.0,5.5) {Angular Spectrum};  
  \node[black] at (8.0,4.5) {\small $  C_\ell \big(z[\chi], z[\chi'] \big)$};
  
  \node[black] at (16.5,5.5) {Observed Power Spectrum};    
  \node[black] at (16.5,4.5) {\small $  P \big( k_{\hat n}, \vec{\ell}/\bar \chi, z[\bar \chi] \big)$};      

\end{tikzpicture}
\caption{Scheme showing the three stages we follow in constructing the observed power spectrum. We start by correlating the 3D density field $\delta(\vec x, z)$, which provides us with the theoretical, unobservable, unequal-time 3D power spectrum $\mathcal P(\vec k, z,z')$. From this, we can construct the observable angular power spectrum $C_\ell(z,z')$. Using the flat-sky approximation we can translate the $C_\ell(z,z')$ into the observable equal-time power spectrum $P(\vec k, \bar z)$ at the mean redshift $\bar z$.}
\label{fig:spectra}
\end{figure*}

In this section, we delineate several different statistical, two-point, observables. We start from the usual theoretical power spectrum (see Table~\ref{tab:notation} and~\cite{Raccanelli+:2023} for details), $\mathcal P (\vec k)$ defined as the ensemble average power spectrum. This power spectrum is not observable as it is properly defined only in a fully 4D space and could be accessible only by a meta-observer outside of the system.

However, from there, we can define a procedure such that by introducing the observer and, therefore, the lightcone for our observations, by taking into account the projections on the sky, we can compute the observable angular spectrum $C_\ell$ (for the derivation of lightcone effects on galaxy clustering, see the pioneering work of~\cite{Matsubara:1999} and e.g.,~\cite{Yoo:2008, Yoo++:2009, Bonvin+:2011, Challinor+:2011, Jeong++:2011, Bertacca:2012, Yoo+:2013} for angular, Fourier, configuration space and spherical-Fourier Bessel).
We note that when defining {\it observable} quantities in this work, we do not include observational effects such as masking and even more instrument-related issues, which go beyond the scope of our work. As we show, this step folds the $\mathcal P (\vec k)$ information coming from the modes along the line of sight into the unequal-time structure of $C_\ell$, thus, effectively performing the compression from 4D into 3D. We construct the observed power spectrum $P_{\rm obs}(\vec k)$ (and we will omit the $obs$ indication from now on), transforming the unequal-time information of  $C_\ell$ and reconstructing the modes along the line of sight. Lastly, we introduce the observable dimensionless frequency-angular power spectrum $\widetilde {\mathbb C}(\omega, \ell)$, which carries information analogous to the observed power spectrum $P(\vec k)$, but free of the assumed fiducial cosmology. We will derive the expression for such observable and argue that it presents several advantages over the standard power spectra used in literature. A summary of the different two-point statistics, both theoretical and observable ones, is given in Table~\ref{tab:notation}.

\subsection{Angular power spectrum} 
\label{sec:angular_spectrum}

The simplest and most natural projected observable we can construct is the angular power spectrum $C_\ell$. For a given 3D density field $\df(\vec r)$, we can introduce the projected statistics using a general window function $W$ as:
\eq{
\hat \df (\vec \theta) 
&= \int d\chi ~ W(\chi) \df \big( \chi \vhat n , \chi \vec \theta, z[\chi] \big) \\
&= \int d\chi ~ W(\chi) \int \frac{d^3 k}{(2\pi)^3}~ 
e^{- i \chi \vec k . ( \vhat n + \vec \theta)} \df \big( \vec k, z[\chi] \big)\, , \non
}
where $\chi$ is the comoving distance, $\chi(z) = c \int_0^z dz~ H^{-1}(z)$, and we use the flat-sky geometric set-up, as depicted in Figure \ref{fig:spectra}. Given that the projected field is limited to a single plane, we can introduce 2D Fourier coordinates:
\eeq{
\hat \df (\vec \ell)  = \int d^2 \theta~ e^{i \vec \ell \cdot \vec \theta} \hat \df (\vec \theta)\, ,
\label{eq:2d_FT}
}
for which we have the corresponding momentum representation:
\eeq{
\hat \df(\vec \ell)   
= \int \frac{d\chi}{\chi^2} ~ W(\chi) \int \frac{d k_{\hat n}}{2\pi}~ e^{- i \chi k_{\hat n}} \df \big( k_{\hat n} \vhat n, \vec{\tilde \ell}, z[\chi] \big)\, ,
}
where we used $\vec{\tilde \ell} = \vec \ell/\chi$. If we consider the two-point correlator, we get:
\eq{
\la \hat \df(\vec \ell) \hat \df^*(\vec \ell') \ra
&= (2\pi)^2 \int \frac{d\chi}{\chi^2}\frac{d\chi'}{\chi'^2} ~ W(\chi) W'(\chi') \label{eq:two_point_function} \\
&\hspace{-1.2cm}\times \df^{\rm 2D} \lb \vec{\tilde \ell} - \vec{\tilde \ell'} \rb  \int \frac{d k_{\hat n}}{2\pi} ~ e^{ i \delta \chi k_{\hat n}}
 \mathcal P \big( k_{\hat n} \vhat n, \vec{k}_\perp, z[\chi], z[\chi'] \big)\, , \non
}
where we introduced the relative radial distance variable $\delta \chi = \chi' - \chi$, and where $\vec{k}_\perp$ is constrained by the relation $\vec{k}_\perp = \vec{ \tilde {\ell}} = \vec{ \tilde {\ell}'}$ (as we are in flat-sky). Given that we have introduced the relative distance variable $\delta \chi$, we should choose the corresponding mean distance $\bar \chi$. We have the freedom to choose the mean distance, and some natural options are:
\eeq{
\bar \chi_{\rm a} \equiv \frac{1}{2} \lb \chi + \chi' \rb, ~~
\bar \chi_{\rm g} \equiv  \sqrt{ \chi \chi' }, ~~
\bar \chi_{\rm h} \equiv \frac{2 \chi \chi'}{\chi + \chi'},
}
which correspond to the arithmetic, geometric and harmonic mean, respectively. Without specifying the choice of the mean distance $\bar \chi$, we can transform the coordinates to obtain:
\eq{
\la \hat \df(\vec \ell) \hat \df^*(\vec \ell') \ra
=& (2\pi)^2
\int d \chi d \chi' \; \frac{\bar \chi^2}{\chi \chi'} 
W\lb \chi \rb W'\lb \chi' \rb \\
&\times \mathcal A(\delta)
\df^{\rm 2D} \big( \vec{\ell} - \vec{\ell}' + \varphi(\delta) \vec \Delta  \big)
\mathbb C \lb \ell, \bar \chi, \delta \chi \rb \, , \non
}
where $\delta = \delta \chi/(2 \bar \chi)$, and $\varphi(\delta)$ is an off-diagonal phase of the Dirac delta function, and we introduced $\bm{\Delta} = \vec{\ell}' + \vec{\ell}$
and the \textit{unequal-time angular power spectrum}:
\eeq{
\mathbb C \lb \ell, \bar \chi, \delta \chi \rb
=  \frac{1}{\chi\chi'} \int \frac{d k_{\hat n}}{2\pi} ~ e^{ i \delta \chi k_{\hat n}}
 \mathcal P \big( k_{\hat n} \vhat n, \vec{k}_\perp, \bar \chi, \delta \chi \big)\, .
}
The factor $\mathcal A(\delta)$ originates from the Dirac delta function and takes a different form depending on the definition of $\bar \chi$. The detailed derivation and a few explicit choices are shown in Appendix~\ref{appA}. 

Since the observer position is fixed, we see that the exact translation invariance in the observer plane does not hold. This is reflected in the fact that the two-dimensional Dirac delta function, besides depending on the wave vectors $\vec \ell$ and $\vec \ell'$, also depends on the distances $\chi$ and $\chi'$. Nonetheless, since we are interested in geometries where the mean distance is much larger than the relative distance $\delta \chi$, it is natural (and it considerably simplifies calculations) to expand around the leading solution that preserves the translation invariance. We can thus express the two-point flat-sky correlators as a sum: 
\eq{
\la \hat \df(\vec \ell) \hat \df^*(\vec \ell') \ra = (2\pi)^2
\df^{\rm 2D} \big( \vec{\ell} - \vec{\ell}' \big) \sum_{n=0}^\infty 
\frac{ \big( \overset{\leftarrow}{\partial}_{\vec \ell'} \cdot \bm{\Delta} \big)^n }{2^n n!} C^{(n)}(\ell)\, ,
}
where the partial derivative in the Taylor expansion acts on the left, producing the derivatives of the delta function, and we introduced the contributing angular spectra:
\eq{
C^{(n)}(\ell) =\int d \chi d \chi' \,
W W'\frac{\bar \chi^2}{\chi \chi'}
\mathcal A(\delta) \varphi(\delta)^n
\mathbb C(\ell, \bar \chi, \delta \chi) \, .
}
Note that the higher derivatives of the Dirac delta function introduce off-diagonal contributions to the angular correlations of $\la \hat \df(\vec \ell) \hat \df(\vec \ell') \ra$, i.e.,~$C^{(n)}(\ell)$ for $n>0$ can be interpreted as a measure of non-diagonal contributions to the usual angular power spectrum. What is the source of these non-diagonal contributions? As we know, in the full-sky treatment, isotropy guarantees the proportionality of the angular power spectrum to the Kronecker delta $\delta^{\rm K}_{\ell \ell'}$, while in the flat-sky approximation, we have obtained this condition from the translational invariance in the plane. However, for two physical modes $\vec k_\perp$ lying on two different redshift planes to agree, we have to readjust the corresponding angles, as stated by the Dirac delta $\df^{\rm 2D} \big( \vec{\ell} - \vec{\ell}' \big)$. This generates the off-diagonal contributions as a consequence of the fixed observer. As expected, the $C^{(n)}(\ell)$ contributions for $n$ values higher than the leading $n=0$, either vanish due to the geometric considerations (e.g.,~if $W=W')$, or are suppressed by the $\varphi(\delta)^n \sim \lb \delta \chi/\chi \rb^n$ factor. In the next section, we focus only on the $n=0$ and simply drop the $n$ order label, i.e.,~we define the flat-sky angular power spectrum:
\eeq{
C(\ell) \equiv C^{(0)}(\ell)\, .
}
The contribution of these higher-order $C^{(n)}(\ell)$ have been numerically investigated in more detail in paper ~\cite{Raccanelli+:2023}, where we show that they tend to be suppressed by at least an order of magnitude on all scales. We note that the proper physical interpretation of these contributions is not as corrections to be added to the flat-sky $C(\ell)$ that would bring it close to the full-sky result. Rather, these should be considered as the error estimates of the flat-sky results, asymptotically approaching the full-sky. 

Before continuing our investigation, let us discuss the options for the choice of $\vec k_\perp$. Firstly, we can constrain our considerations to the scalar case $k_\perp$, as is given by the isotropy in the plane perpendicular to the line of sight. This also holds when redshift space distortions are included in the 3D power spectrum. Moreover, from the Dirac delta function constraint $k_\perp = \tilde \ell = \tilde \ell'$ we again have the freedom to construct our choice for $k_\perp$.  This choice determines at which order in $\delta$ the corrections in $\mathbb C \lb \ell, \bar \chi, \delta \chi \rb$ appear. Choosing $k_\perp = \ell/\chi$ or $k_\perp = \ell/\chi'$ is thus suboptimal as it leads to the $\delta$ corrections we saw above. What is the alternative? We can again choose one among the arithmetic, geometric and harmonic combinations:  
\eeq{
k_\perp = \frac{\ell}{2} \lb \frac{1}{\chi} + \frac{1}{\chi'} \rb \, , ~~ 
k_\perp = \frac{\ell}{\sqrt{\chi \chi'}} \, ,~~
k_\perp = \frac{2 \ell}{\chi + \chi'} \, ,
}
which all provide corrections that are of order $\delta^2$. These have to be again evaluated in the chosen $\bar \chi$, $\delta \chi$ coordinates. For concreteness, if we choose the arithmetic mean $\bar \chi_{\rm a}$, this gives us $\mathcal A (\delta)=\chi^2_{\rm a} (1-\delta^2)^2$, and $\varphi(\delta)=\delta)$. Choosing the geometric definition gives $k_\perp = \ell/(\bar \chi_{\rm a} \sqrt{1-\delta^2})$.
However, if we choose the harmonic $k_\perp = \ell/\bar \chi_{\rm a}$, introducing a shorthand notation:
\eq{
\label{eq:curly_flat}
\overline{\mathbb C} \lb \ell, \bar \chi_{\rm a}, \delta \chi \rb
&= \frac{\bar \chi^2}{\chi \chi'} \mathcal A(\delta)
\mathbb C \lb \ell, \bar \chi_{\rm a}, \delta \chi \rb \\
&=  \frac{1}{ \bar \chi_{\rm a}^2} \int \frac{d k_{\hat n}}{2\pi} ~ e^{ i \delta \chi k_{\hat n}}
 \mathcal P \lb k_{\hat n}, \ell/\bar \chi_{\rm a}, \bar \chi_{\rm a}, \delta \chi \rb\, , \non
}
thus eliminating the dependence on the $\delta$ parameter, besides the explicit dependence in the unequal-time power spectrum. In the rest of the paper, unless otherwise specified, we will adopt this geometry choice and drop the index on the $\bar \chi$ and overline on $\overline{\mathbb C}$. 

The full-sky version of the unequal-time angular power spectrum is well known and given by (neglecting projections effects and window functions):
\eeq{
\label{eq:cell_master_full_sky}
\mathbb{C}^{\rm full}_{\ell}  (\chi_1,\chi_2) \equiv 4\pi \int \frac{k^2 dk}{2\pi^2}\; \mathcal P(k;\, \chi_1,\chi_2) \, j_\ell(k \chi_1) j_\ell(k \chi_2) \, .
}
We shall refer back to this full-sky angular power spectrum results when generalizing the 3D power spectrum from the flat-sky approximation to the full-sky case.

\subsection{Unequal-time power spectrum} 
\label{sec:un-time_3D_PS}

The theoretical power spectrum $\mathcal P$ is an object that we construct out of the ensemble average of density fields given in the 3D space hypersurface at given time-slices and positions. We can define the cross-correlation of two such density fields, not necessarily at the same hypersurface; this gives us the unequal-time, theoretical, power spectrum:
\eeq{
\la \delta (\vec k,z) \delta (\vec k',z') \ra
= (2\pi)^3 \df^{\rm D}(\vec k + \vec k') \mathcal P (\vec k, z, z')\, ,
}
where $z$ is the redshift and $\vec k$ is the Fourier wave mode corresponding to the 3D position vector $\vec x$. This power spectrum is obtained by cross-correlating the 3D density field at two different times of the evolution of the ensemble, as shown on the left-hand side of Figure \ref{fig:spectra}. It is accessible, e.g.,~by {\it meta}-observers and in N-body simulations and theoretical calculations, but not as a real observable, given that every realistic observer has access only to a lightcone-projected subset of information, thus being limited to specific observables, constructed from measured quantities.  

\begin{figure*}[t!]
\centering
\begin{tikzpicture}[scale=.9,every node/.style={minimum size=1cm},on grid]

  \node[black] at (0.5,3.25) {\small Constructed observed power spectrum: $P \big( q_{\hat n}, \ell/\bar \chi, z[\bar \chi] \big)$};     

  \coordinate (c1) at (-8,0);
  \draw[fill=gray!40!white, draw=black, fill opacity = 0.50]
  ($(c1) + (-10:10.12)$) arc (-10:10:10.12)
  --
  ($(c1) + (10:15.075)$) arc (10:-10:15.075)
  -- cycle;
                  
  \draw[fill=gray!80!black, draw=black, fill opacity = 0.45] (2,-1.5) rectangle (7,1.5);  

  \draw[->, line width=0.2mm] (-4.5,0) -- (-3,0);
  \draw[dotted, line width=0.3mm] (-2.95,0) -- (0.5,0);
  \draw[->, line width=0.2mm] (0.5,0) -- (4.5,0);
  \draw[->, line width=0.1mm] (0.5,0.9) -- (2.8,0.9);
  \draw[->, line width=0.1mm] (0.5,-0.9) -- (6.2,-0.9); 
  \draw [dotted, line width=0.3mm] (-2.95,0) to [out=5,in=180] (0.5,0.9);
  \draw [dotted, line width=0.3mm] (-2.95,0) to [out=-5,in=180] (0.5,-0.9);

  \draw[<->, dashed, line width=0.15mm] (2.8,-0.25) -- (6.2,-0.25);
  \draw[dashed, line width=0.15mm] (2.8,-1.5) -- (2.8,1.5);  
  \draw[dashed, line width=0.15mm] (6.2,-1.5) -- (6.2,1.5);  
  \draw[dashed, line width=0.15mm] (4.5,-1.5) -- (4.5,1.5);    

  \node[black] at (-4.5,0.0) {\small \textbullet}; 

  \node[black] at (-4.8,-0.3) {\small $\mathcal O$};  

  \node[black] at (0.85,0.3) {\small $\bar \chi$};
  \node[black] at (0.95,1.2) {\small $ \chi'$};
  \node[black] at (0.85,-0.5) {\small $ \chi$};    
  \node[black] at (5.2,0.0) {\small $\delta \chi$};
  \node[black] at (6.65,1.2) {\small $\mathcal V$};

  \node[black] at (-3.75,0.35) {\small $\hat n$};

   \draw[snake it]  (2.2,-2) -- (6.8,-2);
   \draw[->, line width=0.1mm] (2.2,-2) -- (2.0,-2);
   \draw[->, line width=0.1mm] (6.8,-2) -- (7.0,-2);   
   \draw[snake it]  (7.5,-1.3) -- (7.5,1.3);   
   \draw[->, line width=0.1mm] (7.5,-1.3) -- (7.5,-1.5);
   \draw[->, line width=0.1mm] (7.5,1.3) -- (7.5,1.5);   

  \node[black] at (4.5,-2.5) {\small $q_{\hat n}\sim 2\pi/\delta \chi$};    
  \node[black] at (8.2,0.0) {\small $\bar \ell/\bar \chi$};    

\end{tikzpicture}
\caption{Construction of the observed 3D power spectrum using the perpendicular and line of sight Fourier modes, $q_{\perp}$ and $q_{\hat n}$. $q_{\perp}$ corresponds to $\ell$ divided by the mean comoving distance, i.e.,~$q_{\perp} = \ell/\bar \chi$, while the along the line of sight mode is the Fourier counterpart of the $\delta \chi$ variable, i.e.,~$q_{\hat n} \sim 2\pi/\delta \chi$.}
\label{fig:flat_sky}
\end{figure*}
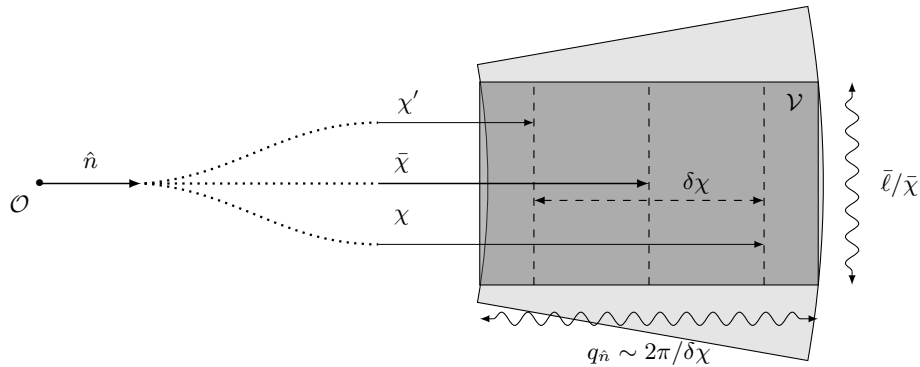

\subsubsection{Tomography with narrow windows} 

Let us look now at expressions for specific choices of narrow window functions. If we introduce infinitely thin redshift slices as $W(\chi) \to \df^{\rm D} \lb \chi - \chi_* \rb$, integrating over the windows gives us $C \big(\ell, \chi_*, \chi'_* \big) = \mathbb C \lb \ell, \chi_*, \chi'_* \rb$, i.e.,~in the limit of narrow window functions we recover the true unequal-time angular power spectrum as an observable. This gives us the relation of the unequal time angular spectrum, measured in the infinitely thin redshift slices, with theoretical power spectrum $\mathcal P(\vec k, z, z')$. Note that we did not specify if our theoretical power spectrum is in real or redshift space since none of the above depends on it. Indeed, we will be able to keep this generality for a while; let us just mention again that in either case, $\mathcal P$ does not depend on the azimuthal angle, which allowed us to drop the angular dependence of the $\vec{k}_\perp$, i.e.,~$\mathcal P(k_{\hat n}, \vec k_\perp, z, z') = \mathcal P(k_{\hat n}, k_\perp, z, z')$. The most common approach in literature at this stage is to resort to the Limber approximation~\cite{Limber:54, Kaiser:1987}. Let us remind ourselves how it can be recovered. We can assume that $\mathcal P$, when integrating over $k_{\hat n}$, predominantly depends on $k_\perp$, neglecting other scale dependencies, i.e.,~$\mathcal P \big( k_{\hat n}, k_\perp, z, z' \big) \simeq \mathcal P \big(k_\perp, z, z' \big)$ and therefore we have:
\eeq{
\mathbb C_{\rm Lim} \lb \ell, \bar \chi, \delta \chi \rb
= \frac{1}{ \bar \chi^2} \delta^{\rm D}(\delta \chi) \mathcal P \big( \ell/ \bar \chi, \bar \chi, \bar \chi \big)\, ,
}
and thus, only the equal-time correlations are not forced to vanish, i.e.,~$\mathbb C (\ell)$ is nonzero only when $z\approx z'$. After restoring the window $W(\chi)$ dependence, we obtain the familiar result:
\begin{equation}
C \big(\ell \big) = \int d \chi \frac{W(\chi)W'(\chi)}{\chi^2} \mathcal P \big( \ell/\chi, \chi, \chi \big)\, .
\end{equation}
On the other hand, if we can assume that the power spectrum $\mathcal P$ has negligible $\delta \chi$ dependence over the relevant integration volume, i.e.,~$z\approx z'$, we can invert Equation~\eqref{eq:curly_flat} to obtain: 
\begin{equation}
\mathcal P \big( k_{\hat n}, \ell/\bar \chi , \bar \chi \big)
=  \bar \chi^2
\int d( \delta \chi)\; e^{-i \delta \chi k_{\hat n}} \mathbb C \big( \ell, \bar \chi, \delta \chi \big)\, .
\end{equation}
In Section \ref{sec:const_3D_PS}, we quantify the errors arising due to this approximation and show the scale-dependent corrections neglected here.

Lastly, let us repeat the statement about the theoretical 3D power spectrum $\mathcal P(\vec k, z, z')$: while it is the quantity that encapsulates the dynamical and stochastic information about the system, it is not observable from the point of view of an observer sitting in some point $\mathcal O$ ``in the box'', as shown in Figure~\ref{fig:spectra} (see also the discussion in~\cite{Raccanelli+:2023}). What observers can measure are unequal-time angular correlations that contain the information imprinted in the ensemble power spectrum, projected into an observable. In the next subsection, we show how the observer can construct the corresponding 3D power spectrum from measurements of the unequal-time angular power spectrum.

\subsection{Observed 3D equal-time power spectrum} 
\label{sec:const_3D_PS}
Our task in this subsection is to \textit{construct the 3D equal time observed power spectrum}, which we call $P(q_{\hat n}, q_\perp, \bar z)$, that an observer at some position $\mathcal O$ could observe. In addition to constructing the power spectrum, we need to define the corresponding Fourier modes $q_{\hat n}$ and $q_\perp$ (which we label with variable $\vec q$ in order to distinguish it from the ensemble power spectrum variable $\vec k$), as well as the mean redshift $z$ (see the set up in Figure~\ref{fig:flat_sky}).
The expectation is that the constructed spectrum is, of course, related to the 3D unequal-time ensemble power spectrum $\mathcal P(\vec k, z, z')$; however, let us proceed step by step in the construction. We use the information from the unequal-time angular power spectra $\mathbb C \lb \ell, \bar \chi, \delta \chi \rb$ in order to construct the Fourier modes along the line of sight $\hat n$, while at the same time keep the information on the mean distance between the observer $\mathcal O$ and the survey volume $\mathcal V$ (or equivalently, the redshift bin analyzed).
Such a fixed volume, depicted in Figure~\ref{fig:flat_sky}, is characterised by the maximal and minimal comoving distances $\chi_{\rm min}$ and $\chi_{\rm max}$. The mean distance $\bar \chi$ can thus take the value between $\chi_{\rm min}$ and $\chi_{\rm max}$, while $ (\chi_{\rm min}-\bar \chi) \leq \delta \chi/2 \leq (\chi_{\rm max} - \bar \chi)$. Sticking to the center of the box with the mean distance $\bar \chi$, to ensure a wide enough range for $\delta \chi$, we can define a constructed wave mode along the line of sight as the Fourier counterpart of $\delta \chi$. We can then use the observable unequal-time angular power spectrum to obtain what we can define as the \textit{observed, equal-time, 3D power spectrum} as:
\begin{equation}
P(q_{\hat n}, \ell/\bar \chi, \bar \chi) \equiv  \bar \chi^2
\int d( \delta \chi )\; e^{-i\delta \chi  q_{\hat n}} \mathbb C( \ell, \bar \chi, \df \chi)\, ,
\label{eq:def_P_obs}
\end{equation}
where $\mathbb C$ is given in Equation~\eqref{eq:curly_flat}. Relying on the set-up shown in Figure~\ref{fig:flat_sky}, we define the perpendicular and line of sight Fourier modes, $q_{\perp}$ and $q_{\hat n}$. With $q_{\perp}$ we identify modes corresponding to $\ell$ divided by the mean comoving distance, i.e.,~$q_{\perp} = \ell/\bar \chi$, while for the line of sight mode we can take the Fourier counterpart of the $\delta \chi$ variable, i.e.,~$q_{\hat n} \sim 2\pi/\delta \chi$. 

When does the, so constructed, observable, 3D power spectrum $P(q_{\hat n}, \ell/\bar \chi, \bar \chi)$ match the theoretical 3D ensemble power spectrum $\mathcal P(\vec k, \chi, \chi)$? As argued in~\cite{Raccanelli+:2023}, this happens when the observed system (the survey) is so small in width and depth compared to the full sky and the distance from the observer that we can approximate the observer as a meta-observer. This here can be compared to assuming negligible dependence of $\mathcal P$ on $\delta \chi$; combining Equations~\eqref{eq:def_P_obs} and~\eqref{eq:curly_flat} we obtain:
\eq{
\bar \chi^2 &\int_{- \delta \chi_<}^{ \delta \chi_>} d( \delta \chi )\; e^{-i\delta \chi  q_{\hat n}} \mathbb C( \ell, \bar \chi, \df \chi) \\
&\hspace{0.3cm}= \int \frac{d k_{\hat n}}{2\pi} ~ \mathcal P \lb k_{\hat n}, \ell/ \bar \chi, \bar z \rb 
\int_{- \delta \chi_<}^{ \delta \chi_>} d( \delta \chi ) ~e^{-i ( q_{\hat n} - k_{\hat n} ) \delta \chi} \non\\
&\hspace{0.3cm}= \int \frac{d k_{\hat n}}{2\pi} ~ \mathcal{B} \lb q_{\hat n} - k_{\hat n}, \delta \chi_<, \delta \chi_> \rb  \mathcal P \lb k_{\hat n}, \ell/ \bar \chi, \bar z \rb\, ,  \non
}
where $\chi_<$ and $\chi_>$ are the closest and farthest integration distances within the survey, and:
\eeq{
\mathcal{B} \lb k, \delta \chi_<, \delta \chi_> \rb 
= \frac{i}{k} \lb e^{-i  \delta \chi_<} - e^{i  \delta \chi_>} \rb\, .
}
Given the dependence of $\mathbb C$ on $\delta \chi$ (that we discuss in Section~\ref{sec:results}), we argue that for large enough survey volumes, the steep decline of $\mathbb C$ from $\delta \chi=0$ values guarantees that we can extend and symmetrize the integration region of $\chi_<$ and $\chi_>$. This is indeed the case for a $\Lambda$CDM universe (see Section~\ref{sec:results}), and thus, as long as $\bar \chi$ is not close to the edge of the survey, we can adapt the limit: 
\eeq{
\mathcal{B} \lb k, \Delta \chi \rb = 2\frac{\sin (k \Delta \chi) }{k} \approx 2\pi \df^{\rm D}(k), ~~ {\rm as}~~ \Delta \chi \to \infty \, ,
} 
where $\chi_< \approx \chi_> \approx \Delta \chi$. Using this approximation, the expression above yields:
\begin{equation}
\label{eq:obs_ps_leading}
P \lb q_{\hat n}, \bar \ell/\bar \chi, \bar z \rb  \approx 
\mathcal P \lb q_{\hat n}, \bar \ell/ \bar \chi, \bar z \rb\, ,
\end{equation}
as expected. The definition given in Equation~\eqref{eq:def_P_obs} thus recovers the original ensemble power spectrum. More generally, the proper 3D observed power spectrum, as defined in Equation~\eqref{eq:def_P_obs}, is sensitive to unequal-time contributions from the 3D theoretical power spectrum $\mathcal P(\vec k, \chi, \chi')$. Before quantifying these unequal-time contributions, we introduce an alternative 3D observable, one that does not rely on the construction of dimensional wave modes $q_{\hat n}$ and $q_\perp$, but still retains the desirable properties of the observed power spectrum $P(q_{\hat n}, q_\perp, \bar z)$.

Let us recap. Cosmological and dynamical information is encapsulated in the 3D unequal-time theoretical power spectrum $\mathcal P(k, z, z')$, and therefore, it is not directly accessible to observations. From this, we can compute observable quantities like unequal time 2D angular power spectra $\mathbb C(\ell, z, z')$. However, the information is spread out (along the line of sight) over different redshift shells. The question is thus whether we can use this angular power spectrum to construct the corresponding equal-time 3D power spectrum that matches the original one as best as possible. The constructed perpendicular modes $q_\perp$ are related to the inverse angular multipoles $\ell$ and the mean distance $\bar \chi$. The modes along the line of sight $q_{\hat n}$ are constructed by Fourier transforming along the unequal-time dependence of $\mathbb C(\ell)$. This construction provides us with the result that (to a very good approximation) corresponds to the familiar equal-time 3D power spectrum. Unequal-time contributions give rise to sub-leading corrections that we quantify in the remainder of the paper.

\begin{figure*}[t!]
\centering
\includegraphics[width=0.95\linewidth]{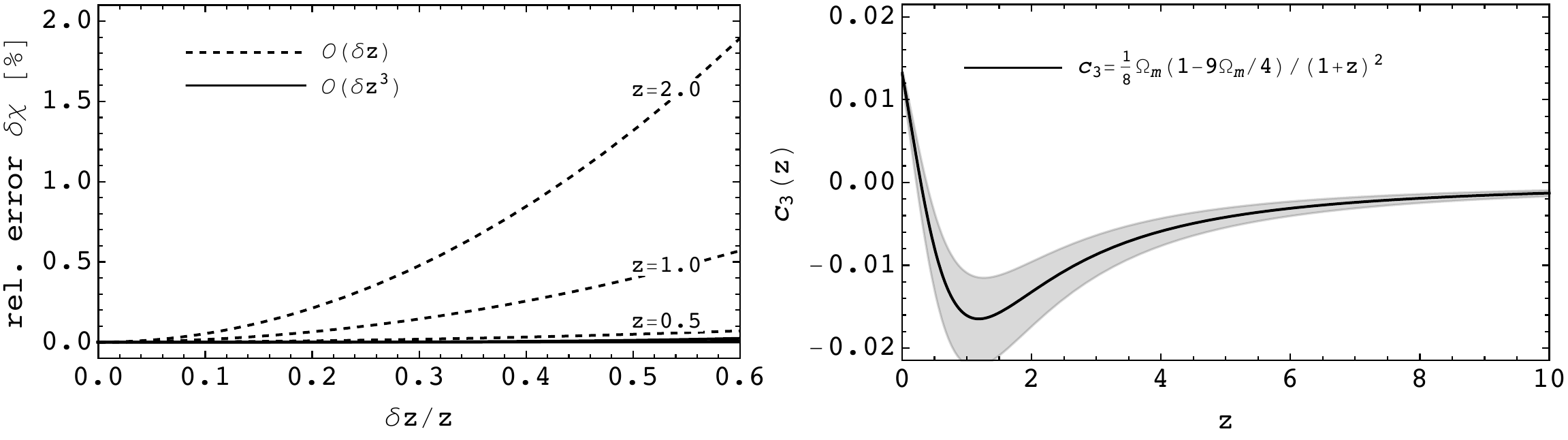}
\caption{Errors on the comoving distance differences, $\delta \chi$. On the left panel, we show the relative errors of the linear approximation $\delta \chi \approx \delta z/ H$ and next-to-linear approximation $\delta \chi \approx \delta z/ H (1- c_3 (\delta z)^2)$ for several mean redshifts $\bar z$, assuming $\Omega_{m0}=0.3$. The right panel shows the size and the redshift dependence of the $c_3(z)$ coefficient for the fiducial value of $\Omega_{m0}$, with the grey band indicating 10\% variations. In all cases, we assume the flat $\Lambda$CDM cosmology.
}
\label{fig:c_3}
\end{figure*}

\subsection{Frequency-angular power spectrum: the ``new observable"}

So far, we have managed to connect the theoretical with the observed power spectrum. As shown, the connection was achieved via the unequal time angular power spectrum $\mathbb C(\ell, z, z')$. Here we revisit the motivation for the construction of the observed power spectrum. We have seen that the information content of the unequal time angular  spectrum is equivalent to the observed power spectrum $P(q_{\hat n}, \ell/\bar \chi, \bar \chi)$, so why bother with the additional step of constructing the observed power spectrum? The reason lies in the compactification of information, i.e.,~the covariance matrix of the observed power spectrum is well described with a diagonal Gaussian approximation. The disadvantage is in the need to construct the observable wave modes $q_{\hat n}$ and $\vec q_\perp$, using the fiducial cosmology to determine the comoving distance. This gives rise to the well-known Alcock-Paczyn\'ski effect~\cite{Alcock+:1979}.

The observed power spectrum is obtained by performing a simple Fourier transform in the $\delta \chi$ variable. However, one can imagine doing the same procedure, as defined in Equation~\eqref{eq:def_P_obs}, without involving the comoving distance, i.e.,~we can define a \textit{frequency-angular power spectrum} as:
\eeq{
\label{eq:tilde_math_C}
\widetilde {\mathbb C}(\omega, \ell, \bar z) \equiv
\int d \delta z ~ e^{- i \omega \delta z} 
\mathbb C(\ell, \chi(\bar z), \delta z)\, ,
}
where the new Fourier frequency variable $\omega$ plays the role previously done by  $q_{\hat n}$. The statistical properties of the covariance matrix should inherit all the properties of the 3D power spectrum (approximate diagonal structure and Gaussianity). We highlight that this new observable, given its functional dependence on only observable quantities, does not exhibit any  Alcock-Paczy\' ski effects, i.e.,~we do not need a fiducial cosmology to compute physical distances, typically required in computing the 3D observed power spectrum. We can generalize this definition even further by introducing the variable frequency $\omega(\ell, \bar z)$, which can also depend on cosmological parameters. We thus obtain a \textit{generalized frequency-angular power spectrum}, defined as:
\eeq{
\label{eq:tilde_math_C_gen}
\widetilde {\mathbb C} (\omega, \ell, \bar z) \equiv
\int d \delta z ~ e^{- i \omega(\ell, \bar z) \delta z} 
\mathbb C(\ell, \chi(\bar z), \delta z)\, .
}
We shall see how to best utilize this generalized form further below.

Later on, we investigate the properties of this observable, assuming some concrete form of the 3D theoretical power spectrum. However, before that, we can again look at the simplifying case when we assumed a negligible unequal-time dependence of the 3D theoretical power spectrum, i.e., we assume $\mathcal P(\vec k, \bar z, \delta z)=\mathcal P(\vec k, \bar z)$. We have:
\eq{
\widetilde {\mathbb C}(\omega, \ell, \bar z) 
&= \int d \delta z ~ e^{- i \omega \delta z} \mathbb C(\ell,\bar \chi(\bar z), \delta z) \\
&=  \frac{1}{\bar \chi^2}  \int \frac{d k_{\hat n}}{2\pi}\; \Omega(\omega, k_{\hat n}) \mathcal P \lb k_{\hat n}, \ell/\bar \chi, \bar z \rb\, ,
}
where:
\eeq{
\Omega(\omega, k_{\hat n}) = \int d \delta z  \; \exp\lb i \delta \chi k_{\hat n} - i \omega \delta z\rb.
}
Assuming linear dependence of $\delta \chi$ on $\delta z$, i.e.,~$\delta \chi \approx \frac{d\chi}{d z}\delta z = \delta z/ H$, where $H$ is the Hubble parameter, evaluated at the mean redshift $\bar z$, we have:
\eeq{
\Omega(\omega, k_{\hat n}) = (2\pi) \delta^{\rm D} \lb k_{\hat n} / H - \omega \rb\, \, ,
}
which gives the frequency-angular power spectrum:
\eeq{
\label{eq:tilde_3D_C}
\widetilde {\mathbb C}(\omega, \ell, \bar z) = \frac{H}{\bar \chi^2} \mathcal P \lb H \omega, \ell/\bar \chi, \bar z \rb\, ,
}
that is a dimensionless quantity, depending only on observable variables. Adding the next order correction to $\delta \chi$, we have $\delta \chi \approx \frac{d\chi}{d z}\delta z = \frac{\delta z}{H} [1- c_3 (\delta z)^2]$, where the expression for the $c_3(\bar z)$ in $\Lambda$CDM is obtained by expanding $\delta \chi(\bar z, \delta z)$ around the equal time case, we have:
\begin{equation}
c_3(z) = \frac{1}{8}\frac{\Omega_{m}(z)\left[ 1 - \tfrac{9}{4} \Omega_{m}(z) \right]}{(1+\bar z)^2 } \; ,
\end{equation}
and the full derivation can be found in Appendix~\ref{appB}.
This gives us:
\eq{
\label{eq:tilde_3D_C_expanded}
\widetilde {\mathbb C}(\omega, \ell, \bar z) 
&= e^{ - c_3 \omega \frac{d^3}{d \omega^3}} \frac{H}{\bar \chi^2}  \mathcal P \lb H \omega, \ell/\bar \chi, \bar z \rb\, \\
&\approx \frac{H}{\bar \chi^2} \lb 1 - c_3 \omega \frac{d^3}{d \omega^3} \rb \mathcal P \lb H \omega, \ell/\bar \chi, \bar z \rb\, . \non
}
In Figure~\ref{fig:c_3}, we show the dependence of the $\delta \chi$ variable on $\delta z$. The purpose of this is to establish the approximations leading to the result in Equation~\eqref{eq:tilde_3D_C}. On the left panel, we show the relative errors of the linear approximation $\delta \chi \approx \delta z/ H$ for several redshifts in $\Lambda$CDM cosmology, where we see that for most low-redshift spectroscopic surveys, the error caused by neglecting higher order $\mathcal O(\delta z^3)$ corrections is suppressed to a fraction of a percent. Moreover, once the $O(\delta z^3)$ corrections are added, corrections are suppressed below the $0.1\%$.

On the right panel, we show the redshift dependence of the $c_3$ coefficient describing the $O(\delta z^3)$ to the $\delta \chi \leftrightarrow \delta z$ relation. The size of the coefficient in a flat $\Lambda$CDM universe depends only on one cosmological parameter, $\Omega_m$, its value being bounded $|c_3|\lesssim0.02$ for values of $\Omega_m$ currently allowed, and it asymptotes to zero as $\sim 1/z^2$. These considerations prompt us to believe that for all practical purposes, in $\Lambda$CDM cosmology, accounting for the leading corrections in $\delta z$ as done in Equation~\eqref{eq:tilde_3D_C_expanded} should suffice for low-redshift surveys, while future surveys that aim to measure galaxy clustering at $z>3$ might need to account for the higher order correction term.

Let us also entertain the fact that the 3D power spectrum $\mathcal P$ is typically a function of $k = \sqrt{k^2_{\hat n} + k_\perp^2}$, and once redshift-space distortions are added, also even powers of the orientation angle $\mu = k_{\hat n}/k$. In terms of our observable quantities, $\omega$ and $\ell$, this means that we have:
\eeq{
k^2 = \frac{1}{{\bar \chi}^2} \left[ \ell^2 + ( H \bar \chi)^2 \omega^2 \right] \, .
}
Motivated by the multipole expansion usually performed in the observed power spectrum, let us look at how we could reproduce this in the case of the frequency-angular power spectrum $\widetilde {\mathbb C}$. First let us define the total momentum $L \equiv \sqrt{ \ell^2 + ( H \bar \chi)^2 \omega^2}$. We can immediately notice that $L$ is no longer a cosmology-independent variable as $\omega$ and $\ell$ are, as it depends on the $H \bar \chi$ product. This is where our generalization introduced in Equation~\eqref{eq:tilde_math_C_gen} becomes useful. We can use the freedom introduced in the generalized frequency $\omega(\ell, \bar z)$ to cancel the cosmology dependence introduced in $k^2$, i.e., we can introduce:
\eeq{
\label{eq:omega}
\omega(\ell, \bar z) \to \frac{1}{ (H \chi)(\bar z)} \omega \, ,
}
which gives us the wave modes:
\eeq{
k^2 = \frac{1}{{\bar \chi}^2} \lb \omega^2 + \ell^2 \rb =  \frac{L^2}{{\bar \chi}^2}  \, ,
}
where the total momentum is simply $L \equiv \sqrt{\omega^2 + \ell^2}$. This allows us to rewrite the frequency-angular power spectrum given in Equation~\eqref{eq:tilde_3D_C} as:
\eeq{
\label{eq:tilde_3D_C_II}
\widetilde {\mathbb C}(L, \bar z) = \frac{H}{\bar \chi^2} \mathcal P \lb L/\bar \chi, \bar z \rb\, .
}
We note that a different choice of $\omega$ than the one given in Equation~\eqref{eq:omega}, or even fixing the fiducial cosmology of the $(H \chi)(\bar z)$ weight, would lead to anisotropic dependencies in $\ell$ and $\omega$, which is equivalent to the Alcock-Paczy\'ski effect \cite{Alcock+:1979}. These anisotropies can then be used to constrain cosmological models; this is also possible since such anisotropies have a different shape dependence than RSD, assuming enough dynamic range is captured.

We can also look at the frequency-angular power spectrum in redshift-space, assuming for now just the usual linear Kaiser formula \cite{Kaiser:1987}. Neglecting, for now, unequal-time effects, we have:
\eeq{
\mathcal P_{\rm lin}(k,\mu,\bar z) = D^2 \left[ 1 + f(\bar z) \mu^2 \right]^2 \mathcal P_{0} (k)\, ,
\label{eq:linear_kaiser}
} 
where we assume that the linear growth factor and rate are evaluated at the mean redshift $\bar z$, i.e.,~$D = D(\bar z)$ and $f = f(\bar z)$. $\mathcal P_{0}$ gives the shape dependence of the linear power spectrum. There is a simple relation between the wave mode angular variable and a newly introduced angular variable $\nu = \omega/L$, that is:
\eeq{
\mu = \frac{k_{\hat n}}{k} = (H \bar \chi)  \frac{\omega(\ell,\bar z)}{L} \to \frac{\omega}{L} = \nu \, .
}
Using the expression given in Equation~\eqref{eq:tilde_3D_C} we obtain the Kaiser frequency-angular power spectrum in redshift-space:
\eeq{
\widetilde {\mathbb C}_{\rm lin}(L, \nu , \bar z) = \frac{H D^2}{\bar \chi^2} \lb 1 + f \nu^2 \rb^2 \mathcal P_{0} (L/\bar \chi)\, .
}
In this observable, the multipoles, obtained by expanding in Legendre polynomials in $\nu$, retain the same form as in the usual 3D observed power spectrum.

\begin{figure*}[htb!]
\centering
\includegraphics[width=0.95\linewidth]{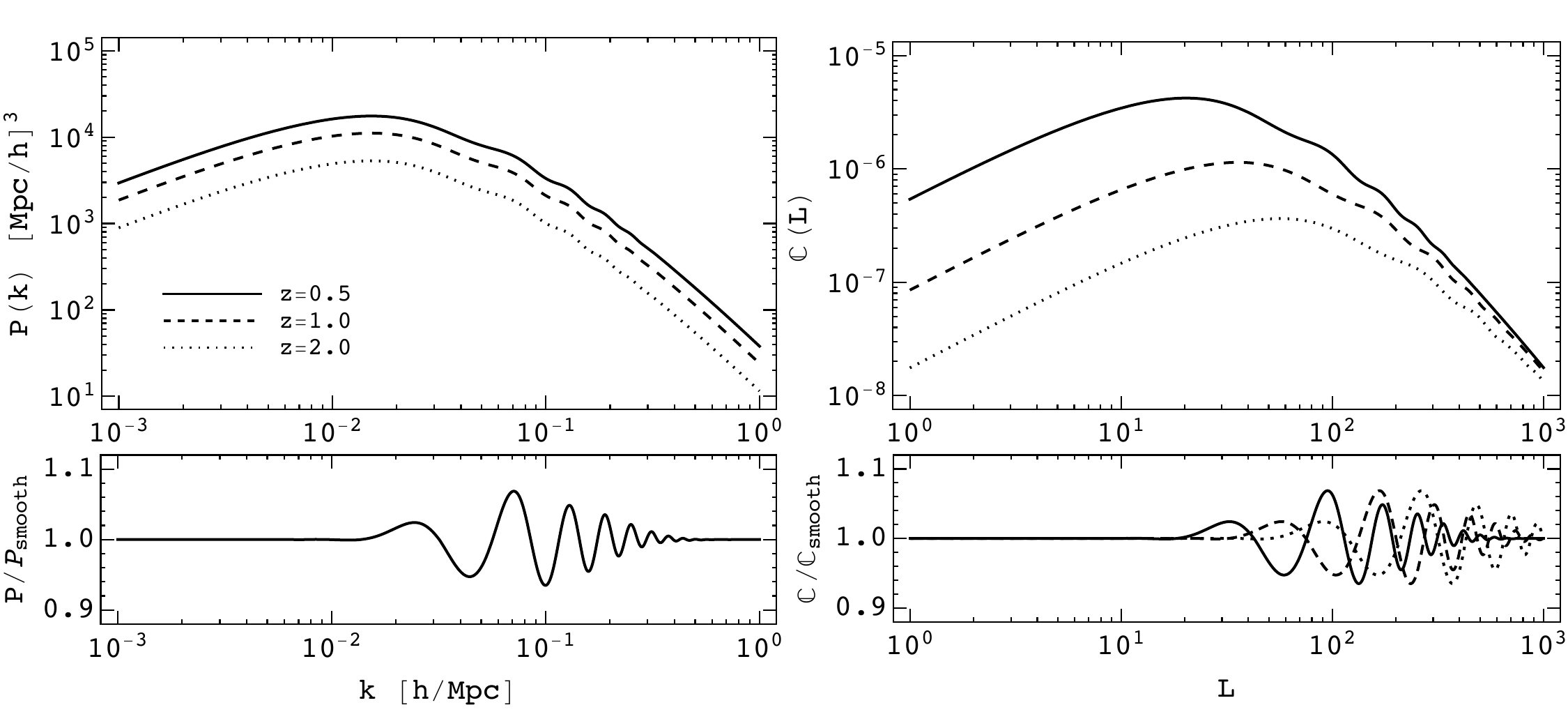}        
\caption{
Linear theory results for the equal-time theory power spectrum (left panels) and the frequency-angular power spectrum $\widetilde {\mathbb C}$ (right panels), for three different redshifts, $z=0.5,~1.0$ and $2.0$. The bottom panels show the ratio with the smooth (no BAO wiggles) version of the respective power spectra. For the equal-time theory power spectrum, the BAO signal is independent of redshift, while for the frequency-angular power spectrum $\widetilde {\mathbb C}$, the signal shifts proportionally to the comoving distance $\chi$.
}
\label{fig:PS_vs_tildeC}
\end{figure*}

In Figure~\ref{fig:PS_vs_tildeC}, we show the comparison of the equal-time linear theory power spectrum and the frequency-angular power spectrum $\widetilde {\mathbb C}(L)$. We show the results for three redshifts $z=0.5,~1.0$, and $2.0$, also showing the ratio with the smooth (no-wiggles) version of the spectrum in the bottom panels. As also indicated in the plots, the usual observed power spectrum $P(k)$ depends on chosen units, while the new power spectrum $\widetilde {\mathbb C}(L)$ is a unit-independent quantity. We also notice that, while in the linear theory power spectrum the BAO signal does not shift with different redshifts, for our new observable the BAO signal shifts with the comoving distance $\chi(z)$. This happens because of the relation between the average comoving distance $\chi$ and the multipole scale. From the bottom panels, which show the ratio to the smooth spectrum, we can see that the features shift from $z_1$ to $z_2$, as $L_2/ \chi_2 = L_1/\chi_1$. This is, of course, similar to the standard Alcock-Paczyn\'ski effect~\cite{Alcock+:1979}, where the anisotropies in the BAO are used to calibrate geometric distances. However, the novelty of the approach taken here is in the relinquishing of the need to use fiducial cosmology and reliance on isotropy. In our formalism, the comoving distance information can be directly established from the position of the BAO wiggles.

We now estimate the corrections to the power spectrum $\widetilde {\mathbb C}(L)$ arising from the $(\delta z)^3$ term in the $\delta \chi \leftrightarrow \delta z$ relation. From Equation~\eqref{eq:tilde_3D_C_expanded}, we see that the leading correction to the power spectrum takes the form:
\eq{
\label{eq:delta_tilde_C}
\delta \widetilde {\mathbb C}_{\rm lin}(\omega, \ell, \bar z)  
&\approx - c_3 \frac{H}{\bar \chi^2} \omega \frac{d^3}{d \omega^3} \mathcal P_{\rm lin} \lb H \omega, \ell/\bar \chi, \bar z \rb \, ,\\
&= - c_3 D^2 H^3 \frac{\nu^2}{L^2} \delta \mathcal P_3 \lb L/\bar \chi, \nu\rb\, , \non
}
where we used the linear matter power spectrum $\mathcal P_{\rm lin}(k,\bar z) = D(\bar z)^2 \mathcal P_{0} (k)$ (neglecting here redshift-space distortions) and we introduced:
\eq{
 \delta \mathcal P_3 \lb k, \mu \rb = \mu^2 & k^3 \mathcal P'''_{0}(k) \\ 
 & + 3 \lb 1 - \mu^2 \rb \left[ k^2 \mathcal P''_{0}(k) - k \mathcal P'_{0}(k) \right] \, . \non
}
The corrections in Equation~\eqref{eq:delta_tilde_C} are proportional to $\nu^2$, therefore introducing an anisotropy even when we start with the isotropic power spectrum $\mathcal P$. This originates in the fact that the introduction of an observer breaks some symmetries in the system, and it shows how the ensemble average power spectrum is not accessible as an observed power spectrum.

Moreover, the corrections depend on the derivatives of the theoretical power spectrum up to the third derivative. In order for the frequency-angular power spectrum $\widetilde {\mathbb C}(L)$ to be useful observable, comparatively to the observed power spectrum $P(k)$ as defined in Equation~\eqref{eq:def_P_obs}, these $\delta \widetilde {\mathbb C}$ corrections should be negligible in all practical cases. We are thus interested in estimating the size of the $\delta \widetilde {\mathbb C}$ corrections. The maximal contribution is expected on large scales, given the $1/L^2$ dependence. Approximating the power spectrum with the power law on large scales, we have $\nu_{\rm max} \approx \sqrt{3/(2(4-n_s))}$, which turns out to be a good approximation on all scales. Using this $\nu_{\rm max}$ value, we can provide an estimate for the corrections to the 
\eq{
\left| \frac{\delta \widetilde {\mathbb C}_{\rm lin}(L, \nu, \bar z)}{ \widetilde {\mathbb C}_{\rm lin}(L, \bar z)} \right| 
&\leq |c_3| \frac{(\bar \chi H)^2}{L^2}\frac{ \delta \mathcal P_3 \lb L/\bar \chi, \nu_{\rm max} \rb}{\mathcal P_0 \lb L/\bar \chi\rb} \, .
}
We can estimate these effects to be of order percent for $L\lesssim 10$ at high redshifts $z \sim 5$, while their size drops quickly for lower redshifts and higher $L$. Their impact thus might be relevant only when considering future wide and deep high redshift surveys.

The conclusion is thus that the frequency-angular power spectrum $\widetilde {\mathbb C}$, as introduced in Equation~\eqref{eq:tilde_math_C_gen}, is a well-behaved observable with small to negligible sub-leading corrections. In that respect, it is equivalent to the observed 3D power spectrum $P(k)$, with the additional advantages already highlighted above.

In this derivation, we adopted the flat-sky approximation and used the corresponding angular power spectrum $\mathbb C_\ell$. However, our newly defined observable, the frequency-angular power spectrum $\widetilde {\mathbb C}$ as defined in Equation~\eqref{eq:tilde_math_C_gen}, does not require a flat-sky approximation. On the contrary, we are free to extend the relationship and introduce the full-sky version of the generalized frequency-angular power spectrum, defined as:
\eeq{
\widetilde {\mathbb C}^{\rm full}_\ell (\omega, \bar z) \equiv \int d \delta z ~ e^{- i \omega(\ell, \bar z) \delta z} \mathbb C^{\rm full}_\ell \left(z_1, z_2\right)\, ,
}
where $\mathbb C_\ell^{\rm full}$ is the full-sky unequal-time angular power spectrum, given in Equation~\eqref{eq:cell_master_full_sky}. Moreover, the observed, equal-time, 3D power spectrum, as introduced in Equation~\eqref{eq:def_P_obs}, can analogously be extended to its full-sky version. We can simply replace the corresponding angular power spectrum $\mathbb C_\ell$:
\eeq{
\label{eq:obs_P_full_sky}
P^{\rm full} \lb q_{\hat n}, \tfrac{\ell}{\bar \chi}, \bar \chi\rb \equiv  \bar \chi^2
\int d( \delta \chi )\; e^{-i\delta \chi  q_{\hat n}} \mathbb C^{\rm full}_\ell( \bar \chi, \df \chi)\, .
}
By adopting this definition for the $P^{\rm full}$, we are abandoning any notion of the construction of the observable 3D power spectrum in a rectangular box and have a procedure to go from the theoretical power spectrum to the observed power spectrum. Besides, this definition naturally incorporates the so-called wide-angle effects, i.e.,~effects arising from deviations from the flat-sky approximation (see, e.g.,~\cite{Szalay++:1997, Papai+:2008, Raccanelli++:2010, Bertacca:2012, Castorina:2017, Pryer:2021}). Quantifying these deviations basically boils down to estimating the difference of using flat-sky vs the full-sky version of unequal-time angular power spectrum $\mathbb C_\ell$ in Equation~\eqref{eq:obs_P_full_sky}. We shall address the quantification of this difference in future work.

\section{Effects of unequal-time cross-correlations} 
\label{sec:results}
In this section, we look at the results of the observed power spectrum, taking into account unequal time effects. To compute them, we start from the unequal time 3D ensemble power spectrum given by linear theory (and the Kaiser formula when including the redshift space distortions). These are chosen as representative of two instructive cases while still being computationally simple. They can also be straightforwardly generalized to include non-linear corrections using canonical perturbation theory approaches (be it in the EFT suit or others, see e.g.,~\cite{Desjacques:2016} for a review). In addition to these two linear theory results, we also consider the linear power spectrum in Lagrangian perturbation theory as a prototypical example of the resummation of the long displacement field contributions \cite{Vlah:2014, Vlah:2015}. Resumming these long displacements, even in the case of the equal time power spectrum, is important since it affects the damping and shape of the BAO oscillations \cite{Senatore:2014, Vlah:2015, Baldauf:2015, Ding:2017}. Moreover, in the case of the unequal time correlators, these displacements are the primary cause of the rapid decorrelation of radial modes and the suppression of unequal time power relative to the equal time correlators (see e.g.,~\cite{Chisari:2019}, and also~\cite{Kitching+:16} for a related discussion). 

\begin{figure*}[t!]
\centering
\includegraphics[width=0.98\linewidth]{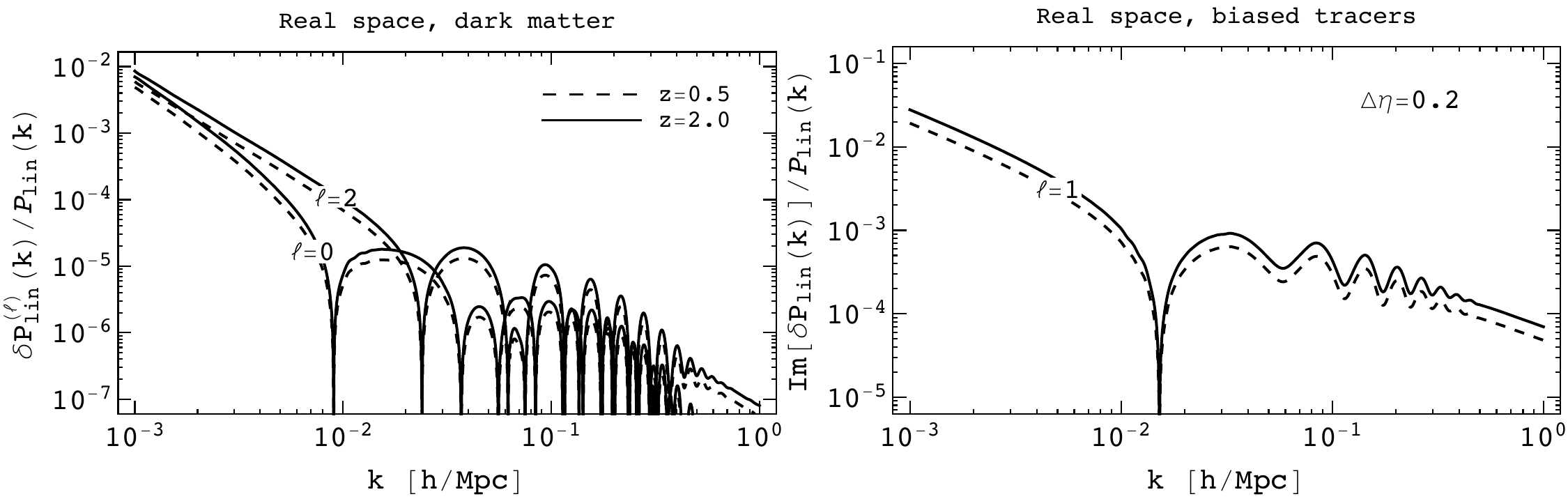}        
\caption{
Linear theory results in a real-space power spectrum for two different redshifts ($z=0.5$, dashed lines, $z=2$, solid lines). {\it Left Panel}: ratio of the correction $\delta P_{\rm lin}$ over the equal time case for the first two even multipoles for real space dark matter spectra. {\it Right Panel}: ratio of the first order imaginary part correction, $\delta P_{\rm lin}$ for $\ell=1$, in a real-space multi-tracer case with $\Delta\eta=0.2$. The ratio is over the amplitude of the (real) monopole, for a comparison of amplitudes.
}
\label{fig:ObsPSRealSpace}
\end{figure*}

Combining Equation~\eqref{eq:def_P_obs}, that defines that observed, equal-time,
3D power spectrum, with the flat-sky approximation for the angular power spectrum given in Equation~\eqref{eq:curly_flat}, we arrive at the following expression:
\begin{equation}
\label{eq:obs_PS_corrections}
P\hspace{-0.1cm}\left(q_{\hat n}, \frac{\ell}{\bar\chi}, \bar \chi\right) \hspace{-0.1cm}
=\hspace{-0.2cm}\int \frac{d k_{\hat n}}{2\pi} d( \delta \chi )
e^{-i\delta \chi  (q_{\hat n} - k_{\hat n})} \mathcal P \hspace{-0.1cm}\lb k_{\hat n}, \frac{\ell}{\bar\chi}, \bar \chi, \delta \chi \rb.
\end{equation}
This expression gives us a direct relationship between the observed equal-time power spectrum $P$ and the theoretical 3D unequal-time power spectrum $\mathcal P$. We note that the unequal-time effects, encapsulated in the $\delta \chi$ power spectrum dependence, are folded in together with the dependence on the modes along the line of sight $k_{\hat n}$. This folding is finally combined in the cumulative (effective) line of sight mode $q_{\hat n}$. In the rest of this section, we investigate the consequences of this folding, investigating the Equation~\eqref{eq:obs_PS_corrections}.

\subsection{Linear Power Spectrum} 
\label{subsec:linPS}

\subsubsection{Dark matter results}

We start our investigation of unequal time effects by first considering just dark matter linear theory results, where the 3D unequal time power spectrum is given by:
\eeq{
\mathcal P_{\rm lin}(k, z_1, z_2) = D(z_1)D(z_2) \mathcal P_{0}(k)\, ,
}
and we can separate the time dependence into two linear growth factors $D(z_i)$ and the time-independent $k$-dependent term. 
In order to proceed, we want to expand around the equal time solution; we follow the analogous procedure as in~\cite{Raccanelli+:2023} and expand the product of the two growth factors up to quadratic order in $\df \chi$ to obtain:
\begin{equation}
\label{eq:corrDM}
D(z_1)D(z_2) = D^2(\bar z) + \frac{1}{8} \Delta^{(0)}_2(\bar z) \left[H(\bar z) \df \chi\right]^2 \, ,
\end{equation}
where we introduced the mean redshift-dependent factor:
\eeq{
\Delta^{(0)}_2 (\bar z)= -2 D^2(\bar z) \left[ 1 + f(\bar z)  -  \frac{3}{2}\frac{\Omega_m(\bar z)}{f(\bar z)} \right] \frac{f(\bar z)}{(1+\bar z)^2} \, ,
}
obtained using the arithmetic definitions for $\bar \chi$ and $\delta \chi$; for the extensive calculation we refer to Appendix~\ref{app:unequal_time}.
It is worth noting that the first order correction vanishes (but as we will see later on, this does not always happen).
Using the expression given in Equation~\eqref{eq:obs_PS_corrections}, we first evaluate the integral over $\delta \chi$ to obtain:
\eq{
\int d( \delta \chi ) \; & e^{-i\delta \chi  (q_{\hat n} - k_{\hat n})} \left[ 1 - \frac{1}{4} \gamma_\times (H \df \chi)^2 \right] \\
&= \left[ 1+ \frac{\gamma_\times}{4} \lb H \partial_{q_{\hat n}} \rb^2 \right]  (2\pi) \delta^{\rm D}  (q_{\hat n} - k_{\hat n}) \, , \non
}
where we use the factor $\gamma_\times = -\frac{1}{2} \Delta^{(0)}_2/D^2$; this gives us the observed power spectrum expression:
\eeq{
P_{\rm lin}(k_{\hat n}, k_\perp, \bar z) = \left[ 1+ \frac{\gamma_\times}{4} H^2 \lb \hat n \cdot \vec \nabla \rb^2 \right]
\mathcal P_{\rm lin} \lb k, \bar z \rb \, .
}
The deviation from the leading result obtained in Equation~\eqref{eq:obs_ps_leading}, i.e.,~the canonical linear theory results, is then given by:
\eq{
\delta P_{\rm lin}(k, \mu, \bar z) 
&= P_{\rm lin}(k, \mu, \bar z) - \mathcal P_{\rm lin} \lb k, \bar z \rb \\
&=  \frac{\gamma_\times}{4}  \lb\frac{H}{k}\rb^2  D(\bar z)^2 \delta \mathcal P_2 \lb k, \mu \rb  \, , \non
}
with:
\eeq{
\label{eq:delta_P_20}
\delta \mathcal P_2 \lb k, \mu \rb = 
 \mu^2 k^2\mathcal P''_{0}(k) + \lb 1 - \mu^2 \rb k \mathcal P'_{0}(k) \,,
}
where $\mu$ is the usual cosine of the angle between the wave mode $\vec k$ and the line of sight, $\mu = k_{\hat n} / k$, and the derivatives $\mathcal P^{(n)}_0$ are to be taken w.r.t. the wave mode $k$. The unequal-time effects can thus give rise to anisotropies in the observed 3D power spectrum, generating higher multipole contributions. Besides contributing to the monopole, Equation~\eqref{eq:delta_P_20} also contributes to the quadruple, and we can write:
\begin{align}
\label{eq:deltaPell}
\delta P^{(0)}_{\rm lin} &=  \frac{\gamma_\times}{12}  \lb\frac{H}{k}\rb^2  D(\bar z)^2 \big( k^2\mathcal P''_0 + 2 k \mathcal P'_0 \big) \, , \\
\delta P^{(2)}_{\rm lin} &=  \frac{\gamma_\times}{6} \lb\frac{H}{k}\rb^2 D(\bar z)^2 \big( k^2\mathcal P''_0 - k \mathcal P'_0 \big) \, . \non
\end{align}{
In the left panel of Figure~\ref{fig:ObsPSRealSpace}, we show the ratio of these corrections for the monopole and quadrupole relative to the linear theory at redshifts $z=0.5$ and $2.0$. It is interesting to notice that even without considering redshift-space distortions, these corrections introduced quadrupole corrections, which is expected as a consequence of breaking the statistical isotropy. In a follow-up work, we intend to compare such corrections to the ones introduced by the Doppler term and relativistic corrections (see, e.g.,~\cite{Raccanelli++:2013, Raccanelli++:2013II, Raccanelli++:2015, Raccanelli++:2016}).

\subsubsection{Multi-tracer analyses}
\label{sec:mt}

Until now, in this section, we considered correlations in real space and for the dark matter case. In the case of biased tracers, the results have to obviously take into account the fact that sources are biased tracers of the underlying matter distribution; however, the structure of the corrections is the same. The result is instead different, as introduced in~\cite{Raccanelli+:2023} when we consider the correlation of two different tracers in the so-called multi-tracer analysis (see e.g.~\cite{McDonald+:2008}).
If we consider the cross-correlation of two different tracers in linear theory, the 3D unequal time power spectrum is given by:
\eeq{
\label{eq:linear_th_bias}
\mathcal P_{\rm lin}^{\rm AB}(k, z_1, z_2) = D(z_1)D(z_2) b_{\rm A}(z_1) b_{\rm B}(z_2) \mathcal P_{0}(k)\, ,
}
where this is the cross-power spectrum of two sources types $\{A,B\}$, at redshifts $\{z_1,z_2\}$ respectively.
Expanding in unequal-time (see Appendix \ref{app:unequal_time}) gives us a non-vanishing linear $\delta \chi$ contribution, unlike the case for single tracer:
\eq{
b_{\rm A}(z_1)b_{\rm B}(z_2)
= \lb 1 + \frac{\Delta_b}{2} H \df \chi \rb b_{\rm A}(\bar z)b_{\rm B}(\bar z)  +  \dots \, ,
}
where we have defined:
\eeq{
\Delta_b = \frac{d}{dz} \ln \lb \frac{b_{\rm A}}{b_{\rm B}} \rb \, .
}
As an example computation of the deviation from the leading order results in Equation~\eqref{eq:obs_PS_corrections}, we can calculate:
\eq{
\int d( \delta \chi ) \; & e^{-i\delta \chi  (q_{\hat n} - k_{\hat n})}  \lb 1 + \frac{1}{2} \Delta_b H \df \chi \rb  \\
&= \lb 1+  \frac{i}{2}  \Delta_b H \partial_{q_{\hat n}} \rb  (2\pi) \delta^{\rm D}  (q_{\hat n} - k_{\hat n}) \, , \non
}
which gives us the observed 3D power spectrum:
\eeq{
P_{\rm lin}(k_{\hat n}, k_\perp, \bar z) = \left[ 1+ \frac{i}{2}  \Delta_b H \left( \hat n \cdot \vec \nabla \right) \right] \mathcal P_{\rm lin} \lb k, \bar z \rb \, .
}
This thus gives rise to an imaginary component of the observable 3D power spectrum. Even though arising from different origins, similar effects are present when gravitational redshift effects are included in the Kaiser formula~\cite{McDonald:2009}. The presence of odd multipoles was also found and discussed in configuration space in~\cite{Raccanelli++:2015}. However, the advantage of this formalism lies in the fact that here multipoles can be calculated using only one Legendre polynomial (as some of the geometrical dependencies are folded during the conversions in Figure~\ref{fig:spectra}), with the introduced error being very small, while in configuration space there are two angles over which we need to integrate.

We can write this first-order deviation from the equal-time case, in the multi-tracer power spectrum as:
\begin{align}
\label{eq:delta_P_lin_chi}
\delta P_{\rm lin}(k, \mu, \bar z) 
&= P_{\rm lin}(k, \mu, \bar z) - \mathcal P_{\rm lin} \lb k, \bar z \rb \\
&= \frac{i}{2}  \Delta_b\,  b_{\rm A}b_{\rm B} \frac{H}{k} \delta \mathcal P_1 \lb k, \mu \rb \, , \non
\end{align}
where:
\begin{equation}
\label{eq:deltaP1}
\delta \mathcal P_1 \lb k, \mu \rb = \mu k \mathcal P'_0 \lb k \rb \, .
\end{equation}
The angular dependence in this term arises from the single derivative along the line of sight and thus gives rise only to a dipole contribution $\delta P^{(1)}_{\rm lin}$. This is an important result, as in the standard calculations, odd multipoles are zero; this could therefore result in a new observable or even a new tool to measure galaxy bias and its evolution~\cite{Spezzati++:2023}.

To continue our investigation, let us focus on a concrete case; we assume that our two tracers have evolution similar to the dark matter but still deviating slightly from each other, i.e.,~we assume a simple model $b = b_{0} D^{\eta}$. We can then write:
\eeq{
\Delta_b = \frac{b'_{\rm A}}{b_{\rm A}}- \frac{b'_{\rm B}}{b_{\rm B}} \approx \frac{(\eta_{\rm B} - \eta_{\rm A})  f} {(1+\bar z)} .
}
This result is also a special case of the expressions given in Appendix \ref{app:unequal_time}.
For concreteness, here we will show results for some examples, and surveys-motivated cases will be presented in a follow-up paper~\cite{Spezzati++:2023}.

We start with a rather conservative case where we assume $\eta_{\rm B} - \eta_{\rm A}= 0.2$, which would correspond to a $\sim10\%$ deviation of the time-evolution of each of the tracers from the dark matter case. On the right panel of Figure~\ref{fig:ObsPSRealSpace}, we show the amplitude of the first-order imaginary correction to the equal time power spectrum. We plot the ratio of $\delta P_{\rm lin}$ of Equation~\eqref{eq:delta_P_lin_chi}, with $\Delta\eta=0.2$. The ratio is over the amplitude of the (real) monopole and should not be intended as an estimate for detection but as a comparison of amplitudes. We will see below that in the case of redshift-space distortions, this generalizes to higher angular contributions, thus giving rise to higher-order odd multipoles.

\subsection{Redshift Space Distortions} 
\label{subsec:RSD}
\begin{figure*}[t!]
\centering
\includegraphics[width=0.98\linewidth]{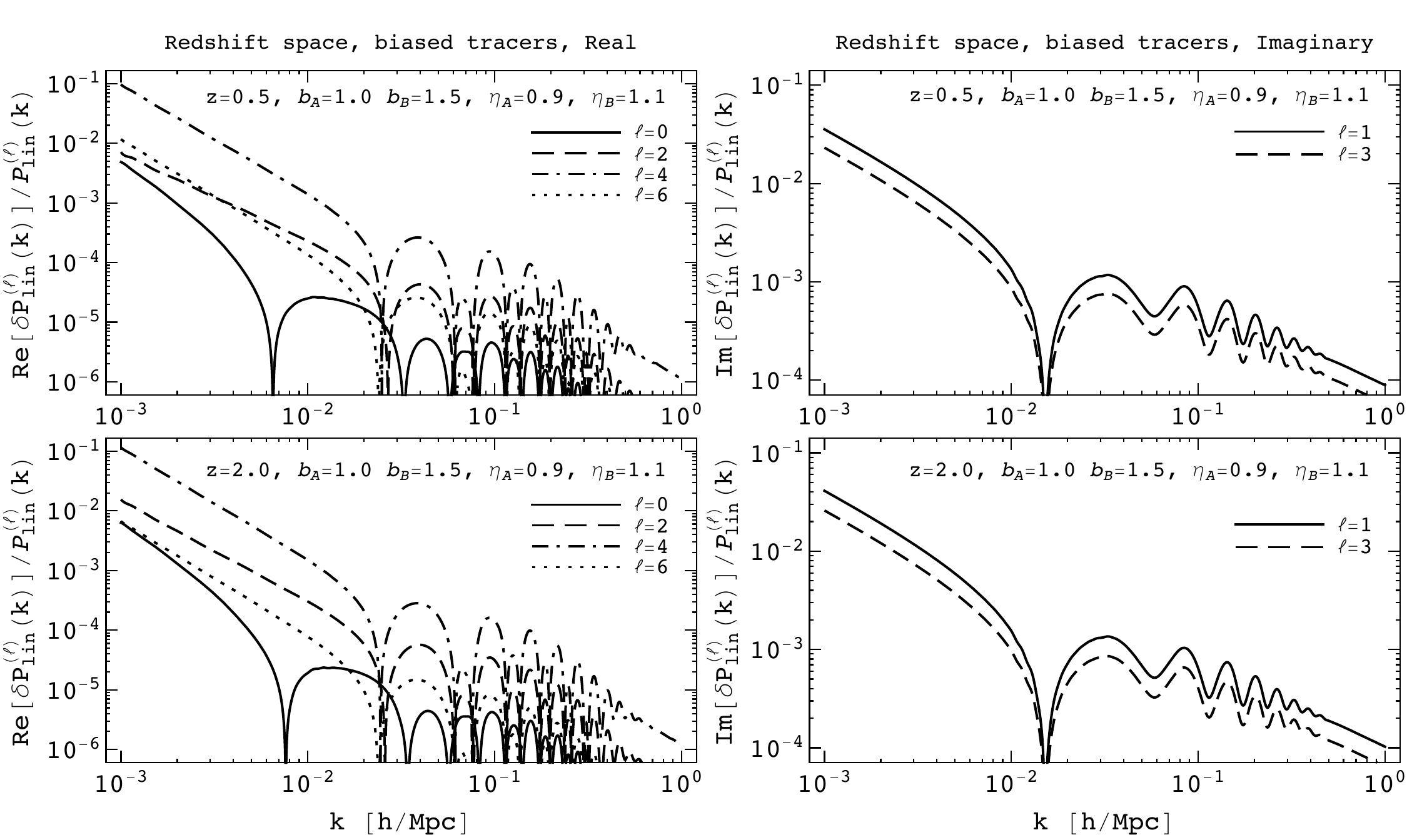}        
\caption{
Linear theory results, as in Figure~\ref{fig:ObsPSRealSpace} in redshift-space for biased tracers, for different multipoles and redshifts, for a particular choice of bias parameters and their time dependence.  Two different redshifts are shown: $z=0.5$ (dashed lines) and $z=2$ (solid lines).
{\it Left Panel}: Even multipoles (real part).
{\it Right Panel}: Odd multipoles (imaginary part). In the case of odd multiples, lines are divided by the linear theory, as in Figure~\ref{fig:ObsPSRealSpace}.
}
\label{fig:ObsPSRedshiftSpacePlot}
\end{figure*}

In this Section, we extend the linear cross-correlation model used in Equation~\eqref{eq:linear_th_bias}, by adding redshift-space distortions. In this work, as a first proof of principle and in the spirit of leaving the theoretical part as simplified as possible in order to have full control of the procedure, we consider just the Kaiser factor~\cite{Kaiser:1987,Hamilton:1997}, where we can replace: 
\begin{equation}
 b_{\rm X}(z_i) \to D(z_i) \left[ b_{\rm X}(z_i) + f(z_i) \mu^2 \right] \; .
\end{equation}
Considering the corrections up to second order in $\delta \chi$, we can expand the product of two Kaiser factors in the unequal time power spectrum as:
\begin{widetext}
\eq{
\lb b_A +fD \mu^2 \rb_{z_1} \lb b_B +fD \mu^2 \rb_{z_2} = \lb b_A +fD \mu^2 \rb \lb b_B +fD \mu^2 \rb 
&+ \frac{1}{2} H \lb \Delta^{(0)}_1 + \Delta^{(1)}_1 f D \mu^2 \rb \delta \chi  \\
& + \frac{1}{8} H^2 \lb \Delta^{(0)}_2 + \Delta^{(1)}_2 f D \mu^2 + \Delta^{(2)}_2 f^2 D^2 \mu^4 \rb \lb \delta \chi \rb^2 + \ldots \, , \non
}
\end{widetext}
where we followed the same procedure of Equation~\eqref{eq:corrDM} for the additional terms. This expression is derived in more detail in Appendix~\ref{app:unequal_time}. After performing the integrals over $\delta \chi$ and $k_{\hat n}$ in Equation~\eqref{eq:obs_PS_corrections}, the deviation from the usual Kaiser observed redshift-space power spectrum can be expressed as follows:
\eeq{
\delta P_{\rm lin}= \frac{i}{2} \frac{H}{k} \Delta_1 \delta \mathcal P_1 - \frac{1}{8}  \lb\frac{H}{k}\rb^2 \Delta_2 \delta \mathcal P_2 \, , 
}
where we have the two factors:
\eq{
\Delta_1 &= \Delta^{(0)}_1 + \Delta^{(1)}_1 f D \mu^2 \, , \\
\Delta_2 &= \Delta^{(0)}_2 + \Delta^{(1)}_2 f D \mu^2 + \Delta^{(2)}_2 f^2 D^2 \mu^4 \, , \non
}
and the $\delta \mathcal P_1$ and $\delta \mathcal P_2$ are the $k$ dependent contributions given in Equations~\eqref{eq:deltaP1},~\eqref{eq:delta_P_20} respectively. The time dependent functions $\Delta^{(i)}_1$ and $\Delta^{(i)}_2$ are derived in Appendix~\eqref{app:unequal_time}.

Collecting together the different contributions, we can therefore express the deviations from the standard equal time power spectrum multipoles as:
\begin{equation}
\label{eq:rsd_PS_corrections}
 \delta P^{(\ell)}_{\rm lin} \hspace{-0.1cm}= \hspace{-0.05cm}i \frac{H}{k} {\rm Im} \hspace{-0.1cm}\left[ \tau^{(1)}_\ell \right] \hspace{-0.1cm}(k \mathcal P'_0) - \lb\frac{H}{k}\rb^2 \hspace{-0.25cm}{\rm Re} \hspace{-0.1cm}\left[ \tau^{(2)}_\ell k^2 \mathcal P''_0 + \tau^{(1)}_\ell k \mathcal P'_0 \right],
\end{equation}
where $\tau_\ell$ are the multipole coefficients, and their expressions are presented in Table~\ref{tab:c_rsd_coeffs}. The real part gives rise to even, and the imaginary part to odd multipoles.

In Figure~\ref{fig:ObsPSRedshiftSpacePlot}, we show results for the case presented above; once again, we leave a detailed estimation of the magnitude of these corrections and their detectability (and/or need to be included in the galaxy clustering modeling) for a dedicated separated study. These results show that unequal-time corrections are generally small; however, there are scenarios in which they could become relevant. At large scales and high-z, they can contribute up to several percent to the total observed power spectrum, making it relevant for planned future surveys that aim to target exactly large cosmological volumes at very high redshifts.

Moreover, such corrections depend on the different ways galaxies trace the underlying matter distribution, introducing anisotropies even without considering redshift-space distortions, and they generate odd multiples. Interestingly, unequal-time corrections depend on the derivative of the power spectrum, which enables the study of its slope and runnings in a novel way. Finally, we note that the presented results rely on a specific choice of bias values and their evolution, and we used the Kaiser formula for redshift-space distortions. We intend to extend this analysis  to include relativistic effects in a hypothetical multi-tracer future analysis in a follow-up paper.

\begin{table*}[t!]
\centering
\begin{tabular}{ l  l}
\hline
\hline
Real part of $\tau^{(m)}_\ell$ &    \\ [3pt]
\hline 
$~~\tau^{(1)}_0  = \frac{1}{12}  \Delta_2^{(0)} + \frac{1}{60} f D \Delta_2^{(1)} + \frac{1}{140} (f D)^2 \Delta_2^{(2)} $ & $~~\tau^{(2)}_0  = \frac{1}{24}  \Delta_2^{(0)} + \frac{1}{40} f D \Delta_2^{(1)} + \frac{1}{56} (f D)^2 \Delta_2^{(2)}$  \\[5pt]
$~~\tau^{(1)}_2  = -\frac{1}{12}  \Delta_2^{(0)} + \frac{1}{84} f D \Delta_2^{(1)} + \frac{1}{84} (f D)^2 \Delta_2^{(2)} $ & $~~\tau^{(2)}_2  = \frac{1}{12}  \Delta_2^{(0)} + \frac{1}{14} f D \Delta_2^{(1)} + \frac{5}{84} (f D)^2 \Delta_2^{(2)}  $ \\[5pt]
$~~\tau^{(1)}_4  = -\frac{1}{35} f D \Delta_2^{(1)} - \frac{4}{385} (f D)^2 \Delta_2^{(2)} $ & $~~\tau^{(2)}_4 = \frac{1}{35} f D \Delta_2^{(1)} + \frac{3}{77} (f D)^2 \Delta_2^{(2)} $ \\[5pt]
$~~\tau^{(1)}_6  = - \frac{2}{231} (f D)^2 \Delta_2^{(2)} $ & $~~\tau^{(2)}_6  =  \frac{2}{231} (f D)^2 \Delta_2^{(2)} $ \\[5pt]
\hline
\hline
Imaginary part of $\tau^{(m)}_\ell$ &   \\ [3pt]
\hline
$~~\tau^{(1)}_1  = \frac{1}{2} \Delta_1^{(0)} +  \frac{3}{10} f D \Delta_1^{(1)}  $ & \\ [5pt]
$~~\tau^{(1)}_3 = \frac{1}{5} f D \Delta_1^{(1)} $ & \\ [5pt]
\hline
\end{tabular}
\caption{Real and imaginary parts of the multipole coefficients $\tau^{(m)}_\ell$ used in the linear redshift-space distortions given in Equation~\eqref{eq:rsd_PS_corrections}. Explicit forms of the $\Delta^{(i)}_1$ and $\Delta^{(i)}_2$ functions are given in Appendix~\eqref{app:unequal_time}.}
\label{tab:c_rsd_coeffs}
\end{table*}

\subsection{IR resummation of Power Spectrum} 
\label{subsec:IR_resumm}

The unequal-time theoretical power spectrum in Lagrangian perturbation theory (LPT) can be schematically written as \cite{Vlah:2014, Vlah:2015}:
\eeq{
\label{eq:LPT_UT_PS}
\mathcal P(k,z,z') = \int d^3q\ e^{i\vec k\cdot\vec q- \frac 1 2 k_ik_j A_{ij}(\vec q, z, z')} \xi^{(0)}(q,\bar z) +\ldots
}
where the two-point displacement cumulant is given by
\eeq{
A_{ij}(\vec q, z, z') = \la \Delta_i \Delta_j \ra_c ,
}
and $\Delta_i = \psi_i(\vec q_2,z') - \psi_i(\vec q_1,z)$. We have also written the leading order two-point correlator $ \xi^{(0)}(q)$, which can take the form of the linear correlation function in the case of IR resummed linear theory or simply unity in the case of the Zeldovich power spectrum. The dots in Equation~\eqref{eq:LPT_UT_PS} give us the higher-order perturbative corrections as well as the higher unequal-time corrections. For a more detailed discussion and derivation of the unequal-time LPT results, we refer the reader to Appendix~\ref{app:UT_LPT}. Here we just note that the unequal-time corrections given as the dotted expansion in $\delta \chi$ lead to $(H/k)$ type of corrections in the observed power spectrum as given in Subsections~\ref{subsec:linPS} and \ref{subsec:RSD}, and we thus rather focus on the unequal-time effects of the two-point displacement cumulant $A_{ij}$. In Appendix~\ref{app:UT_LPT} we show that we can expand the second displacement cumulant around the mean redshift as
\eeq{ 
\label{eq:A_expansion}
A_{ij}(\vec q,z,z')  \approx  A_{ij}(\vec q,\bar z) + \delta A_{ij}(\vec q,\bar z) (H \df \chi)^2\, .
}
Consequently, the observed power spectrum, given by the expression in Equation~\eqref{eq:obs_PS_corrections}, is
\eq{
P(q_{\hat n}, \ell/\bar \chi, \bar \chi) 
&= \int \frac{d k_{\hat n}}{2\pi} d^3q\; e^{i\vec k\cdot\vec q} e^{- \frac 1 2 k_ik_j A_{ij}(\vec q, z)} \\
&\hspace{-1cm} \times \int d( \delta \chi )  e^{-i\delta \chi  (q_{\hat n} - k_{\hat n})}  e^{- \frac 1 2 k_ik_j \delta A_{ij}(\vec q, z)  (H \df \chi)^2 },  \non
}
where multiplicative terms like $\xi^{(0)}$ can be easily added into consideration, as they do not affect the $\delta \chi$ integral; similarly, the higher $\delta \chi$ terms arising from the unequal-time part of the $\xi^{(0)}$-like operators can be added, as shown earlier in this section.

Using the quadratic expansion of the  displacement cumulant  $A_{ij}$ given in Equation~\eqref{eq:A_expansion}, the $\delta \chi$ integral can be done analytically. We obtain the Gaussian integral form
\eq{
\int d( \delta \chi)\; & e^{-i\delta \chi  (q_{\hat n} - k_{\hat n})}  e^{- \frac 1 2 k_ik_j \delta A_{ij}(\vec q, z)  (H \df \chi)^2 } \\
&= \frac{\sqrt{2\pi}}{H \sqrt{| k_ik_j \delta A_{ij} |}} \exp \lb - \frac{1}{H^2} \frac{(q_{\hat n} - k_{\hat n})^2}{2 k_ik_j \delta A_{ij}} \rb \non \; .
}
If we assume that $\delta A_{ij}(\vec q, z)$ is approximately scale-independent, as argued in Appendix~\ref{app:UT_LPT}, we can write $\delta A_{ij}(\vec q, z) \simeq 2 \delta \sigma(\bar z)^2 \delta^K_{ij} $, where $\delta\sigma^2$ can be interpreted as the average long displacement dispersion due to the unequal-time effects. Estimated lower and upper bounds of $\delta\sigma^2$ values are in Equation~\eqref{eq:sigma_bound}. Using this approximation, we have 
\eq{
P(q_{\hat n}, \ell/\bar \chi, \bar \chi) 
&\simeq \int \frac{d k_{\hat n}}{2\pi} d( \delta \chi )\;  e^{-i\delta \chi  (q_{\hat n} - k_{\hat n})} \\
&~~ \times e^{-k^2 \delta \sigma^2  (H \df \chi)^2}
\int d^3q\; e^{i\vec k\cdot\vec q} e^{- \frac 1 2 k_ik_j A_{ij}(\vec q, z)} \non\\
&= \frac{1}{H \delta \sigma}  \int \frac{d k_{\hat n}}{2 \sqrt \pi k}\; e^{-\frac{(q_{\hat n} - k_{\hat n})^2}{4 k^2 (H  \delta \sigma)^2}}
\mathcal P \lb k_{\hat n}, \ell/\bar \chi, \bar \chi \rb\, . \non
}
This result tells us that the unequal time effects of long displacement modes on the observed power spectrum is to smear the theoretical 3D power spectrum on scales corresponding to $\sim k H \delta \sigma$. For parameters of the $\Lambda$CDM cosmology that we are using here, $2H \delta \sigma$ peaks at $z\sim 0.55$, achieving values of $2H \delta \sigma\sim 0.001$. This provides us with the smoothing kernel of width smaller than any feature in the 3D power spectrum of the $\Lambda$CDM universe, i.e. we can treat $\mathcal P$ effectively as a constant over the integration region where the exponential function has support. Moreover, in the integrand, we can approximate $k = \sqrt{k_{\hat n}^2 + k^2_\perp}\approx  \sqrt{q_{\hat n}^2 + k^2_\perp}$, 
which makes the integral of a simple Gaussian form. Consequently, these simplifications give us  
\eeq{
P(q_{\hat n}, \ell/\bar \chi, \bar \chi) 
\approx \mathcal P \lb q_{\hat n}, \ell/\bar \chi, \bar \chi \rb \, ,
}
i.e. we can neglect the unequal time effects due to the IR resummation. This picture changes if we want to discuss the deviation from the $\Lambda$CDM model, where the power spectrum would exhibit some additional features on scales $k_*\lesssim k H \delta \sigma$. This has an immediate consequence for cosmological models predicting the linear power spectrum with ``features'' - either imprinted during inflation or induced by non-standard expansion histories (see, e.g. \cite{Chluba:2015, Slosar:2019, Achucarro:2022} for recent reviews). Current results suggest that future surveys will be able to detect or tightly constrain features in the primordial spectrum below the one percent level across a wide range of scales \cite{Chen:2020,Ferraro:2022}. This is a far larger effect than the limit imposed due to the long displacement smearing we are considering here.

\section{Conclusion} 
\label{sec:conclusion}

In this paper, we develop a framework for observables of galaxy clustering; in particular, we investigate the role of unequal-time effects in the observed power spectrum $P$. Namely, when constructing the observed power spectrum, we use different redshift slice information to construct the modes along the line of sight. However, the 3D theoretical power spectrum $\mathcal P$ of different redshift slices is inevitably described as an unequal-time power spectrum. This implies that these unequal-time effects and the modes along the line of sight are folded on top of each other in the observed power spectrum. We thus first delineate the connection between the observed equal-time power spectrum and the theoretical 3D unequal-time power spectrum. This connection is accomplished by relying on the flat-sky approximation of the unequal-time angular power spectrum $\mathbb C_\ell$. 
 
In this construction process, we show that one is free to consider also alternative 3D statistics to the canonically defined observed power spectrum $P$. We thus construct an observable frequency-angular power spectrum $\widetilde {\mathbb C}$ and show how this newly introduced statistic naturally includes radial mode contributions and how we can eliminate the need for a priori distance measure assumptions, usually needed in the wave mode construction (so-called Alcock-Paczyn\'ski effects.) This enables one to make measurements independently of the choice of a cosmological model, by introducing a dimensionless quantity depending only on observable variables (Fourier counterparts to the angles and redshifts). We also investigate the properties of this new statistic and verify that, in most current practical applications, it retains all beneficial properties of the canonical observed power spectrum $P$ and that the residual contributions to the modes along the line of sight generated by the redshift dependence can be safely disregarded. Another powerful aspect of this frequency-angular power spectrum $\widetilde {\mathbb C}$ is that the BAO features shift with cosmological distance, making it possible to infer distances directly by using the position of BAO peaks, which is a robust and well-understood measurement.

In the latter part of the paper, we focus on a formulation of the Fourier-space $P(k)$ that includes corrections due to unequal-time when correlating sources (or bins of them) at different redshifts. Starting from the observable angular spectrum, we show how to calculate contributions along the line of sight and quantify them for some example cases. Starting from the equal-time standard case, we find an expression for a series expansion to include unequal-time terms and calculate their amplitude and scale dependencies. Such corrections generally appear at second order in the radial separation between sources, $\delta \chi$. Still, there will be a contribution from the first order when cross-correlating sources with a different bias. These first-order terms give rise to an imaginary part of the power spectrum, which translates into odd multipoles when performing the classic Legendre polynomials expansion. Moreover, unequal-time corrections generate higher-order multipoles, including the odd ones, even in the RSD case (where typically only even multipoles appear). This might represent a new cosmological observable with a yet unexplored potential. We note that such contributions, originating from observable projection effects, are expected to appear also in the higher n-point functions, with the consequence of giving rise to contributions that might be expected to be zero from purely theoretical considerations.

We find that unequal-time corrections give rise to terms typically scaling with $\mathcal H/k$. These contributions are generally small, but they present some interesting features. First of all, multi-tracer analyses depend on the difference between the tracer biases but also on their time derivatives, introducing the exciting possibility of studying the bias evolution in a new way. In redshift space, this dependence extends to derivatives of the growth rate, again opening up a new possible avenue for studying cosmological models. 

As a last part, we consider unequal-time effects arising due to the long displacement field via the IR resummation mechanism. We model these contributions at the linear level of Lagrangian perturbation theory re-summing the linear displacements. We show that unequal-time contributions result in effective smoothing of the original equal-time power spectrum on scales $k_* \sim k H \delta \sigma$ (with $\delta \sigma$ of order few Mpc$/h$). The cumulative effect is thus far smaller than what can potentially be probed by current and upcoming experiments.

In summary, we investigated the effects of the unequal-time contributions in the observed power spectrum and some representative case studies based on examples of the source biases and their redshift evolution. We defined a new observable in angular-frequency space that naturally includes transverse and radial modes and promises to become a more convenient way to analyze galaxy surveys than the canonical observed power spectrum. In Fourier space, our calculation of unequal time effects unveiled a deeper understanding of the behaviour of galaxy clustering along the line of sight, which opens up the possibility of adding a new tool for cosmological studies with galaxy clustering measurements.

\begin{acknowledgments}
AR acknowledges funding from the Italian Ministry of University and Research (MIUR) through the ``Dipartimenti di eccellenza'' project ``Science of the Universe''.
Z.V. is partially supported by the Kavli Foundation.
\end{acknowledgments}  

\appendix

\onecolumngrid
\section{Angular power spectrum and the choice of the mean distance }
\label{appA}

Keeping in mind the different options for the choice of the $\bar \chi$,
starting from Equation~\eqref{eq:two_point_function} the angular power spectrum 
can thus be written as 
\eq{
\la \hat \df(\vec \ell) \hat \df(\vec \ell') \ra
= (2\pi)^2 \int \frac{d \chi}{\chi^2} \frac{d \chi'}{\chi'^2} \, W\lb \chi \rb W'\lb \chi' \rb
\df^D \lb \vec{\tilde \ell} + \vec{\tilde \ell'} \rb
\int \frac{d k_{\hat n}}{2\pi} ~ e^{ i \delta \chi k_{\hat n}}
 \mathcal P \big( k_{\hat n} \vhat n, \vec{k}_\perp, \chi, \chi' \big)\, . 
}
Using the delta function representation in the new variables, we can write
\eeq{
\df^{\rm 2D} \lb \vec{\tilde \ell} + \vec{\tilde \ell'} \rb =
\df^{\rm 2D} \lb \frac{\chi' \vec{\ell} + \chi \vec{\ell'}}{\chi\chi'} \rb = \bar \chi^2 \mathcal A(\delta)\,  \df^{\rm 2D} \lb \vec{\ell} + \vec{\ell'} + \varphi(\delta) \vec \Delta \rb \,,
}
where $\vec \Delta = \vec{\ell}' - \vec{\ell}$, $\bar \chi$ in the mean distance,  $\delta = \frac{1}{2}\delta \chi/\bar \chi$, and $\varphi(\delta)$ is an off-diagonal phase of the Dirac delta function. Specifically, for arithmetic, geometric and harmonic coordinates, respectively, this gives us
\eq{
\df^{\rm 2D} \lb \vec{ \tilde \ell} + \vec{\tilde \ell}' \rb 
&=  \df^{\rm 2D} \lb \frac{\vec{\ell} + \vec{\ell}' - \delta \vec \Delta}{\chi_{\rm a} (1-\delta^2)} \rb
= \chi^2_{\rm a} (1-\delta^2)^2 \df^{\rm 2D} \lb \vec{\ell} + \vec{\ell}' - \vec \Delta \delta \rb \, , \\
\df^{\rm 2D} \lb \vec{ \tilde \ell} + \vec{\tilde \ell}' \rb 
&= \df^{\rm 2D} \lb \frac{(\vec{\ell} + \vec{\ell}')\sqrt{1+\delta^2} - \delta \vec \Delta }{\chi_{\rm g}} \rb
=  \frac{\chi^2_{\rm g}}{(1+\delta^2)} \df^{\rm 2D} \lb \vec{\ell} + \vec{\ell}' - \vec \Delta \delta/\sqrt{1+\delta^2}  \rb \, , \non\\
\df^{\rm 2D} \lb \vec{ \tilde \ell} + \vec{\tilde \ell}' \rb 
&= \df^{\rm 2D} \lb \frac{(\vec{\ell} + \vec{\ell}')\lb 1 + \sqrt{1+4\delta^2} \rb - 2 \delta \vec \Delta }{\chi_{\rm h} \lb 1 + \sqrt{1+4\delta^2} \rb } \rb
=  \chi^2_{\rm h} \df^{\rm 2D} \lb \vec{\ell} + \vec{\ell}' - 2 \vec \Delta \delta/\big(1 + \sqrt{1+4\delta^2} \big) \rb \, , \non
}
where we can identify the factor $\mathcal A$ and phase $\varphi$ in each case.
We have
\eq{
\la \hat \df(\vec \ell) \hat \df(\vec \ell') \ra
= (2\pi)^2
\int d \chi d \chi'\; 
W\lb \chi \rb W'\lb \chi' \rb
\frac{\bar \chi^2}{\chi \chi'} \mathcal A(\delta)
\df^{\rm 2D} \big( \vec{\ell} + \vec{\ell}' + \varphi(\delta) \vec \Delta  \big)
\mathbb C_\ell \lb \chi, \chi' \rb \, ,
}
where we have
\eeq{
\mathbb C_\ell \lb \chi, \chi' \rb
= \frac{1}{\chi \chi'} \int \frac{d k_{\hat n}}{2\pi} ~ e^{ i \delta \chi k_{\hat n}}
 \mathcal P \big( k_{\hat n} \vhat n, \vec{k}_\perp, \chi, \chi' \big)\, .
}
Since we can write 
\eeq{
\df^{\rm 2D} \big( \vec{\ell} + \vec{\ell}' + \varphi(\delta) \vec \Delta  \big)
= \df^{\rm 2D} \big( \vec{\ell} + \vec{\ell}' \big) + 
\Big( e^{\varphi(\delta) \vec \Delta \cdot  \overset{\rightarrow}{\partial}_{\vec \ell}} - 1 \Big) \df^{\rm 2D} \big( \vec{\ell} + \vec{\ell}' \big) \, ,
}
we have
\eeq{
\la \hat \df(\vec \ell) \hat \df(\vec \ell') \ra
= (2\pi)^2
\df^{\rm 2D} \big( \vec{\ell} + \vec{\ell}' \big) \sum_{n=0}^\infty 
\frac{ ( \overset{\leftarrow}{\partial}_{\vec \ell'} \cdot \vec \Delta )^n }{n!} C^{(n)}(\ell)\, ,
}
and where we have introduced
\eeq{
C^{(n)}(\ell) = \int d \chi d \chi'\; 
W\lb \chi \rb W'\lb \chi' \rb
\frac{\bar \chi^2}{\chi \chi'} \mathcal A(\delta) \varphi(\delta)^n
\mathbb C(\ell,\chi, \chi')\, . 
}

\section{From redshift to comoving distance}
\label{appB}

Let us also expand $\delta \chi(\bar z, \delta z) = \chi(\bar z + 1/2\delta z) - \chi(\bar z - 1/2\delta z)$ as a function of $\delta z$, we have:
\eeq{
\delta \chi = \frac{d \chi(\bar z)}{d \bar z} \delta z + \frac{1}{3} \frac{d^3 \chi(\bar z)}{d \bar z^3} (\delta z/2)^3 + \ldots 
= \frac{1}{H(\bar z)} \delta z + \frac{1}{24} \lb \frac{d^2}{d \bar z^2} \frac{1}{H(\bar z)} \rb \delta z^3 + \ldots \, .
}
Since we can write:
\eeq{
\frac{d^2}{d z^2} \frac{1}{H(z)} 
= -  3 \frac{\Omega_{m}\lb 1 - \tfrac{9}{4} \Omega_{m}  \rb}{(1+z)^2 H(z) }\, ,
}
this gives us:
\eeq{
\label{eq:delta_chi_expand}
\delta \chi 
= \frac{1}{H(\bar z)} \delta z - \frac{1}{8}\frac{\Omega_{m}\lb 1 - \tfrac{9}{4} \Omega_{m}  \rb}{(1+\bar z)^2 H(\bar z) } \delta z^3 + \ldots 
\simeq  \left[ 1 - c_3 (\delta z)^2 \right] \frac{\delta z}{H(\bar z)}\, ,
}
and thus:
\eeq{
\Omega(\omega, k_{\hat n}) \equiv \int d \delta z  \; e^{ i \delta \chi k_{\hat n} - i \omega \delta z} 
= e^{ - c_3 \omega \frac{d^3}{d \omega^3}}  (2\pi) \delta^{\rm D} \lb \frac{k_{\hat n}}{H} - \omega \rb
\simeq \lb 1 - c_3 \omega \frac{d^3}{d \omega^3} \rb  (2\pi) \delta^{\rm D} \lb \frac{k_{\hat n}}{H} - \omega \rb \, .
}

\section{Unequal time contributions to the Kaiser terms}
\label{app:unequal_time}

Here we derive the expansion up to the second order in the unequal time variable around a mean.  We treat the deviation $\df \chi$ around the mean comoving distance as the small contribution. We subsequently check the validity of this expansion on several examples. A redshift-dependent physical quantity $F$ we can then simply expand as:
\eeq{
F(z[\chi_i(\chi,\df \chi)]) = D(z[\chi_i(\chi,0)]) 
+\frac{d}{d \df \chi} F(z[\chi_i(\chi,\df \chi)]) \Big|_{\df \chi=0} \df \chi 
+\frac{1}{2} \frac{d^2}{d^2 \df \chi} F(z[\chi_i(\chi,\df \chi)]) \Big|_{\df \chi=0} (\df \chi)^2 + \dots\, , 
}
where $i\in\{1,2\}$ labels the two positions we are concerned with when considering two-point correlations. For the first derivative, we have:
\eeq{
\frac{d}{d \df \chi} F(z[\chi_i(\chi,\df \chi)])
= \frac{d \chi_i}{d \delta \chi}  \frac{dz}{d \chi_i} \frac{d}{dz} F(z[\chi_i(\chi,\df \chi)]) 
= H F' \frac{d \chi_i}{d \delta \chi}\, ,
}
where we use the label $F' \equiv dF/dz$, and also have $dz/ d \chi_i = (d \chi_i /d z)^{-1} = H$. For the second derivative, we have:
\eeq{
\frac{d^2}{d^2 \df \chi} F(z[\chi_i(\chi,\df \chi)])
= \frac{d}{d \df \chi} \lb H F' \frac{d \chi_i}{d \delta \chi} \rb 
= H \lb H F' \rb' \lb \frac{d \chi_i}{d \delta \chi} \rb^2 + H F' \frac{d^2 \chi_i}{d \delta \chi^2} \, . 
}
Using the arithmetic coordinate setup, i.e. the coordinates defined relative to the arithmetic mean, we have $d\chi_1/d \delta \chi = 1/2$, $d\chi_2/d \delta \chi = -1/2$, and thus:
\eeq{
\frac{d}{d \df \chi} F(z[\chi_{1/2}(\chi,\df \chi)])
= \pm \frac{1}{2} H F'\, , ~~{\rm and}~~ 
\frac{d^2}{d^2 \df \chi} F(z[\chi_i(\chi,\df \chi)])
= \frac{1}{4} H \lb H F' \rb' = \frac{1}{4} H^2 \Big(  (\ln H)' F' + F'' \Big) \, .
}
In the case of redshift-space distortions, we need to evaluate the factor:
\eq{
\label{eq:Kaise_UT_corrections}
\lb b_A +fD \mu^2 \rb_{z_1} \lb b_B +fD \mu^2 \rb_{z_2}
&= \lb b_A +fD \mu^2 \rb \lb b_B +fD \mu^2 \rb + \frac{1}{2} H \lb \Delta^{(0)}_1 + \Delta^{(1)}_1 f D \mu^2 \rb \delta \chi  \\
&\hspace{4.6cm} + \frac{1}{8} H^2 \lb \Delta^{(0)}_2 + \Delta^{(1)}_2 f D \mu^2 + \Delta^{(2)}_2 f^2 D^2 \mu^4 \rb \lb \delta \chi \rb^2 + \ldots \, , \non
}
where it is convenient to introduce the factors that depend on the mean redshift:
\eq{
\label{eq:Delta_n}
\Delta^{(0)}_1 &= b'_A b_B - b'_B b_A \, , ~~
\Delta^{(1)}_1 = b'_A -b'_B + \gamma_1(b_B - b_A) \, , \\
\Delta^{(0)}_2 &= b_A  \lb  \gamma_0 b_B' + b_B'' \rb +  \lb  \gamma_0 b_A' + b_A'' \rb b_B - 2 b'_A b'_B\, , ~
\Delta^{(1)}_2 = \gamma_2 b_A +  (\gamma_0 - 2\gamma_1) b_A' + b_A'' + A\leftrightarrow B\, , ~
\Delta^{(2)}_2 = 2 \lb \gamma_2 - \gamma^2_1 \rb \, . \non
}
Above, we introduced factors $\gamma_0$, $\gamma_1$ and $\gamma_2$ that are functions of mean redshift. They are introduced by taking the redshift derivatives of the Hubble parameter, linear growth $D$ and its logarithmic growth rate $f$. Starting from the Hubble parameter, we introduced $\gamma_0 \equiv (\ln H)' = \frac{3}{2} \frac{\Omega_{m}}{(1+z)}$, while the redshift derivative of  linear growth is simply $D'= - f D/(1+z)$. Equation of motion for the growth rate $a d f/d a = - f(2+f) + (1+f)\frac{3}{2} \Omega_m$, gives us:
\eeq{
(\ln f)' =  \frac{1}{1+z} \lb 2+f - (1+f)\frac{3}{2} \frac{\Omega_m}{f} \rb =  \gamma_1 + f/(1+z) ,
}
where we introduced the factor $\gamma_1 \equiv  \lb 2 - (1+f)\frac{3}{2} \frac{\Omega_m}{f} \rb/(1+z)$ and thus
$(f D)' = \gamma_1 f D$. We also have:
\eq{
 \gamma_2 \equiv  \frac{1}{f H D} \lb H (fD)'\rb'  =  \gamma'_1 + \gamma_0 \gamma_1+\gamma_1^2 \, ,
}
where, using $(\Omega_m)' = 3 \Omega_m \Omega_\Lambda/(1+z)$, we can compute the redshift derivative of $\gamma_1$, to obtain:
\eeq{
 \gamma'_1  = \lb \Big( 1 + (1+z)\gamma_1/f - 3 (1+f) \Omega_\Lambda/f \Big) \gamma_0 - \gamma_1 \rb  \frac{1}{1+z}\, .
}
For general tracers, their biases and their time evolution can differ from tracer to tracer and which consequently gives rise to linear corrections in $\delta \chi$ in Equation~\eqref{eq:Kaise_UT_corrections}. We see that $\Delta^{(0)}_1$ contributions can be generated either by different bias values or different bias change rates, while in the case of the redshift-space related $\Delta^{(1)}_1$ term, we have the bias values, and bias change rates contribute additively. On the other hand, second-order contributions are also present in the auto-correlations of any tracer. 

In the case of dark matter, i.e. $b_A=b_B=D$, these expressions can be significantly simplified, as we shall see below. However, in the case of dark matter, it is useful to identify the specific factors of the powers of $\mu^2$ in the Kaiser formula, as it can serve as the crosscheck of the validity of our expansion, i.e. if the expansion up to the second order in the $\delta \chi$ suffices. In this case, as in the case of all autocorrelations, the first-order contributions vanish ($\Delta^{(0)}_1=\Delta^{(1)}_1=0$), while in the case of the second-order contributions, we can identify:
\eq{
\label{eq:producst_comp}
\frac{D_1D_2}{D^2} - 1 \approx \frac{ \Delta^{(0)}_2}{D^2} \frac{1}{8} \lb H \delta \chi \rb^2\, , ~~
\frac{\lb f_1 + f_2 \rb}{2 f} \frac{D_1D_2}{D^2}  - 1 \approx \frac{\Delta^{(1)}_2}{2D} \frac{1}{8} \lb H \delta \chi \rb^2\, , ~~
\frac{f_1f_2}{ f^2} \frac{D_1D_2}{D^2} - 1 \approx \Delta^{(2)}_2 \frac{1}{8} \lb H \delta \chi \rb^2 \, .
}
To proceed a bit further, let us assume a simple power-law model for the bias time dependence $b = b_{0} D^{\eta}$. This gives $b' = \eta b (\ln D)'$ and $b'' = \eta b \lb (\ln D)'' + \eta ((\ln D)')^2 \rb$, and thus besides the first derivative we have stated above, we also need the second derivative $D'' = \lb - f D/(1+z) \rb' =  \big( 1 - (1+z)\gamma_1 \big) \frac{f D}{(1+z)^2}$. Combining and using these in Equation~\eqref{eq:Delta_n}, we obtain for the linear terms in $\delta \chi$:
\eeq{
\label{eq:Deltas_1}
\Delta^{(0)}_1 =  \frac{(\eta_B - \eta_A)}{1+z} f b_Ab_B\, , ~~~
\Delta^{(1)}_1 =  (\eta_B b_B - \eta_A b_A) \frac{f}{1+z} + \gamma_1(b_B - b_A) \, , 
}
both of which vanish in case $b_A=b_B=D$, as we have stated above. For the second order $\delta \chi$ contributions we have:
\eq{
\label{eq:Deltas_2}
\Delta^{(0)}_2 &= b_A b_B \lb - (\eta_A + \eta_B)  \lb 1 + f  -  \frac{3}{2}\frac{\Omega_m}{f}  \rb  + (\eta_A - \eta_B)^2 f \rb \frac{f}{(1+z)^2}    \, , \\
\Delta^{(1)}_2 &= b_A \lb \gamma_2 + (\gamma_1 - \gamma_0) \frac{\eta_A f}{1+z}  + \lb 1 + (\eta_A-1)f \rb \frac{\eta_A f}{(1+z)^2} \rb  + A\leftrightarrow B\, , \non\\
\Delta^{(2)}_2 &= 2 \lb \gamma_2 - \gamma^2_1 \rb \, . \non
}
In the case of dark matter, when $b_A=b_B=D$, we have:
\eq{
\label{eq:Deltas_2_2}
\Delta^{(0)}_2 &= -2 D^2 \lb 1 + f  -  \frac{3}{2}\frac{\Omega_m}{f} \rb \frac{f}{(1+z)^2}    \, , \\
\Delta^{(1)}_2 &= 2 D \lb \gamma_2 + (\gamma_1 - \gamma_0) \frac{f}{1+z}  + \frac{f}{(1+z)^2} \rb \, , \non\\
\Delta^{(2)}_2 &= 2 \lb \gamma'_1 + \gamma_0 \gamma_1 \rb \, . \non
}
These expressions can be used to check the ones given in Equations~\eqref{eq:producst_comp}. In Figure \ref{fig:delta_chi_sqr}, we compare these relations. The points are obtained from direct calculations given on the right-hand side of Equations~\eqref{eq:producst_comp}, while solid lines represent the quadratic approximations whose coefficients are given by Equations~\eqref{eq:Deltas_2_2}. We see that the agreement between the two is excellent in all three cases, and thus we can conclude that the expansion up to the quadratic order suffices for estimating the unequal time effects in any current galaxy survey. 

\begin{figure}[t!]
\centering
\includegraphics[width=0.98\linewidth]{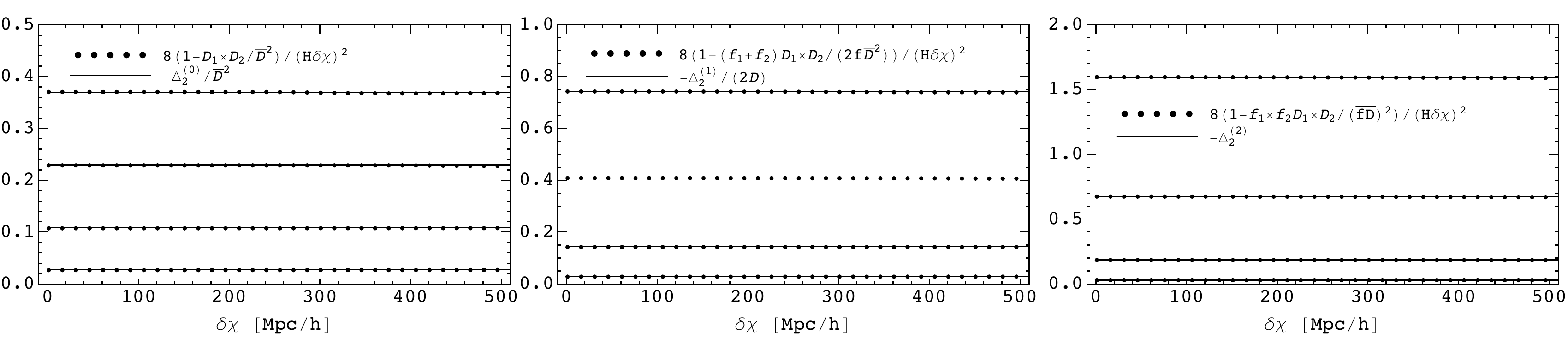}
\caption{
Comparison of the unequal-time Kaiser terms. Three panels show the three unequal-time contributions in the linear Kaiser power spectrum proportional to $\mu^0$, $\mu^2$ and $\mu^4$. These come as different combinations of the linear growth factor $D(z)$ and growth rate $f(z)$ shown as two-point unequal-time factors that appear in the power spectrum, given as a function of comoving distance $\delta \chi$. Points represent the direct calculations, while solid lines give the values of the coefficients up to the $\delta \chi^2$ order. We see that in all cases, the agreement with the quadratic approximation is excellent.
}
\label{fig:delta_chi_sqr}
\end{figure}

\section{Unequal-time power spectrum in Lagrangian perturbation theory}
\label{app:UT_LPT}

In the Lagrangian formalism for describing gravitational clustering, the theoretical power spectrum of a general biased tracer field can be expressed as
\eeq{
\mathcal P(\vec k,z,z') =  \int d^3q~e^{- i \vec k \cdot \vec q} \lb \sum_{\rm a,b} c_{\rm a} c_{\rm b} \la O_{\rm a}(\vec q_1, z) O_{\rm b} (\vec q_2, z') e^{ i \vec k \cdot ( \psi (\vec q_2,z') - \psi (\vec q_1,z) ) } \ra - 1 \rb \, ,
}
where $\psi(\vec q)$ is the displacement field, relating the Lagrangian particle position $\vec q$ to the Eulerian position $\vec x = \vec q + \psi(\vec q)$, and $O_{\rm a}(\vec q, \tau)$ and $c_{\rm a}$ are the set of operators and corresponding biased coefficients describing specific biased tracers. Given our interest in the unequal-time effects in the observable two-point statistics, we are interested in estimating the corrections around some mean redshift $\bar z$. The product of the two bias operators $OO'$ is not of particular interest, given that it also leads to the $(\mathcal H/k)$ type of correction we have investigated in Appendix~\ref{app:unequal_time}, and the results of which we have estimated in earlier parts of this section. Here we shall thus focus on the effects of the long displacement components $\psi(\vec q)$. 

Without going into the details (see \cite{Vlah:2014, Vlah:2015, Vlah:2018, Chen:2020_II} for some recent work on Lagrangian perturbation theory), we can represent the theoretical two-point function
\eq{
\mathcal P(k,z,z') = \int d^3q\ e^{i\vec k\cdot\vec q} \xi^{(0)}(q, \bar z) e^{- \frac 1 2 k_ik_j A_{ij}(\vec q, z, z')} + \ldots
  ~~~{\rm where}~~~ 
  A_{ij}(\vec q, z, z') = \la \Delta_i \Delta_j \ra_c ,
}
$\Delta_i = \psi_i(\vec q_2,z') - \psi_i(\vec q_1,z)$ is the difference of the linear displacements, and the $\ldots$ represents the higher order perturbative terms, as well as the unequal-time expansion term in powers of $(H\delta \chi)$, following the procedure given in Appendix \ref{app:unequal_time}. If, for example, we consider dark matter dynamics, $ \xi^{(0)}(q)$ can be interpreted as a linear correlation function, and the first $\mathcal P(k)$ term above is simply the IR-resummed linear power spectrum (in the equal-time limit). The second displacement cumulant $A_{ij}$ can be decomposed as follows:
\eq{
  A_{ij}(\vec q,z,z') 
  &= \delta^{\rm K}_{ij} X(q,z,z') + \hat{q}_i\hat{q}_j Y(q,z,z') \\
  &= \frac{1}{3}\delta^{\rm K}_{ij} \lb D^2 + D'^2 \rb \Xi_{0}(0)
  - \frac{2}{3}\delta^{\rm K}_{ij} D D' \Xi_{0}(q)
  + 2\left(\hat{q}_i\hat{q}_j-\frac{1}{3}\delta^{\rm K}_{ij}\right) D D' \Xi_{2}(q)\, , \non
}
where we have introduced $\Xi_{0}(q) = \int_0^\infty \frac{dk}{2\pi^2}~ \mathcal P_0 (k) j_0(kq)$ and $\Xi_{2}(q) = \int_0^\infty \frac{dk}{2\pi^2}~ \mathcal P_0(k) j_2(kq)$, and where the scale-dependent part of the linear power spectrum $\mathcal P_0$ is equivalent to the one introduced in Equation~\eqref{eq:linear_kaiser}.
Equivalently, we can write
\eq{
X(q,z,z') &= \frac{1}{3} \lb D^2 + D'^2 \rb \Xi_{0}(0) - \frac{2}{3} DD'\lb \Xi_{0}(q) + \Xi_{2}(q) \rb\, , \\ 
Y(q,z,z') &= 2 ~DD'\Xi_{2}(q)\, . \non
}

\begin{figure}[t!]
\centering
\includegraphics[width=0.98\linewidth]{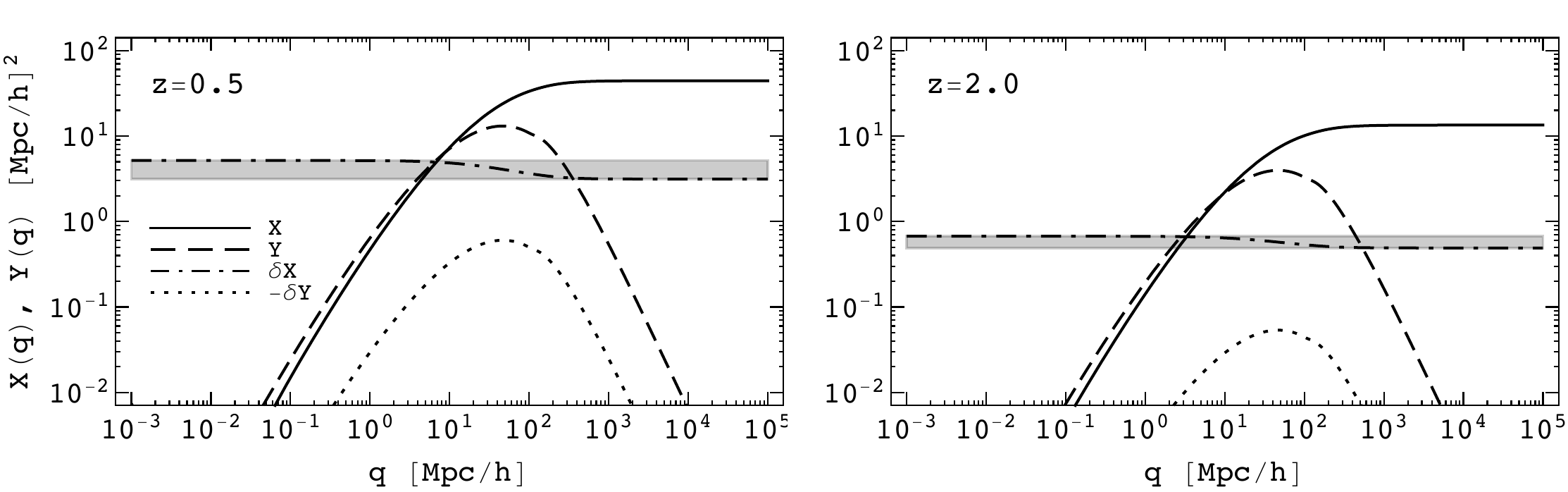}
\caption{
Scale dependence of the equal-time displacement correlators $X$ and $Y$ and corresponding unequal-time contributions $\delta X$ and $\delta Y$ are shown for two different redshifts, $z=0.5$ and $2.0$.  $\delta Y$ contributions are suppressed relative to the equal-time counterpart $Y$ by at least an order of magnitude on all scales and for all redshifts. Conversely,  $\delta X$ exhibits approximately constant behaviour in $q$, with the grey band indicating the range between the to limiting regimes $\delta X(q\to 0)$ and $\delta X(q\to \infty)$. 
}
\label{fig:deltaA}
\end{figure}

If we consider only the unequal-time Zeldovich power spectrum limit, these results are equivalent to the ones obtained in \cite{Chisari:2019}. The unequal-time product of two growth rates we can expand up to the second order in $\delta \chi$, in a similar way done in Appendix \ref{app:unequal_time}, giving us
\eeq{
\frac{D(z_1)D(z_2)}{D(\bar z)^2} \approx 1 - \frac{1}{4} \gamma_\times (H \df \chi)^2\, ,
}
where we introduced the factor $\gamma_\times = f \lb 1 + f - \frac{3}{2}\frac{\Omega_m}{f} \rb /(1+z)^2$. Besides the product, we are also interested in expanding the sum of the two growth factors. This gives us 
\eeq{
\frac{D(z_1)^2 + D(z_2)^2}{2 D(\bar z)^2} \approx 1 - \frac{1}{4} \gamma_+  (H \df \chi)^2\, ,
}
where we introduced the factor $\gamma_+ = f \lb 1 - f - \frac{3}{2}\frac{\Omega_m}{f} \rb /(1+z)^2$.
Using this expansion, we can express 
\eq{
X(q,z,z') &= X(q,\bar z) + \delta X(q,\bar z) (H \df \chi)^2 \, , \\
Y(q,z,z') &= Y(q,\bar z) + \delta Y(q,\bar z) (H \df \chi)^2 \, , \non
}
where we introduced the quadratic corrections 
\eq{
\delta X(q) &= \frac{1}{6} D^2 \bigg(\gamma_\times \left( \Xi_{0}(q) + \Xi_{2}(q) \right) - \gamma_+ \Xi_{0}(0) \bigg)\, , \\ 
\delta Y(q) &= -\frac{1}{2} D^2 \gamma_\times \Xi_{2}(q)\, . \non
}
Accordingly, we can thus write $ A_{ij}(\vec q,z,z')  =  A_{ij}(\vec q,\bar z) + \delta A_{ij}(\vec q,\bar z) (H \df \chi)^2$. In Figure~\ref{fig:deltaA} we show the magnitudes and scale dependence of the $X$ and $Y$ correlators, as well as their corresponding unequal-time contributions $\delta X$ and $\delta Y$ for two different redshifts, $z=0.5$ and $2.0$. We are interested in the relative behaviour of the $\delta X$ and $\delta Y$ compared to the $X$ and $Y$. First, we notice that $\delta Y$ contributions are more than an order of magnitude smaller than $Y$ on all scales at low redshifts, with this difference further decreasing at higher redshifts, i.e. $\delta Y/Y\lesssim1/10$ on all scales and for all redshifts. We can thus neglect $\delta Y$ contributions from our further considerations. For $\delta X$, similar argumentation is not valid, as it is the contribution that dominates on small scales at any redshift. The scale dependence of $\delta X$ is actually bounded for at any $z$ so that we can write $0.3 \lesssim (\delta X(q\to0) - \delta X(q\to\infty))/\delta X(q_{\rm BAO}) \lesssim 1.3$ for all $z$, where the lower bound is reached for high $z$ and vice versa. In other words, the grey bound shown in Figure~\ref{fig:deltaA}, which indicates the deviation of $\delta X$ from some constant value, gets narrow at higher redshifts. Combining these considerations justifies the following approximation for the unequal-time contribution of the displacement correlator
\eeq{
\delta A_{ij}(\vec q,\bar z)  \simeq \delta X(q,\bar z) \delta^{\rm K}_{ij} 
\simeq \frac{1}{6} D(\bar z)^2 \Big( \gamma_\times(\bar z) \Xi_{0}(q) - \gamma_+(\bar z) \Xi_{0}(0) \Big) \delta^{\rm K}_{ij}
\simeq 2 \delta \sigma (\bar z)^2 \delta^{\rm K}_{ij} \, ,
}
where $\delta\sigma^2$ is the scale-independent displacement dispersion due to the unequal-time effects. As discussed above, the magnitude of 
$\delta\sigma^2$ is bounded from above and below, and we can thus write 
\eeq{
\label{eq:sigma_bound}
- \frac{1}{12} \gamma_+(\bar z) \leq \frac{\delta\sigma(\bar z)^2}{D(\bar z)^2  \Xi_{0}(0)} \leq \frac{1}{12}\lb \gamma_\times(\bar z) - \gamma_+(\bar z) \rb \, .
}
We rely on this approximation in deriving the estimates of the unequal-time effects due to the IR-resummation of long displacements discussed in Subsection \ref{subsec:IR_resumm}.

\section*{References}
\bibliography{ms}

\end{document}